# Shaping Quantum Photonic States Using Free Electrons

A. Ben Hayun[1], O. Reinhardt[1], J. Nemirovsky[1], A. Karnieli[1], N. Rivera[2], and *I. Kaminer[1]

[1] Department of Electrical Engineering and Solid-State Institute, Technion, Israel Institute of Technology, 32000 Haifa, Israel.

[2] Department of Physics, Massachusetts Institute of Technology, Cambridge, MA 02139.

**It is a long-standing goal to generate robust deterministic states of light with unique quantum properties, such as squeezing, sub-Poissonian statistics and entanglement. It is of interest to consider whether such quantum states of light could be generated by exploiting interactions with free electrons, going beyond their already ubiquitous use in generating classical light. This question is motivated by developments in electron microscopy, which present a new platform for manipulating photons through their interaction with quantum free electrons. Here, we explore the shaping of photon statistics using the quantum interactions of free electrons with photons in optical cavities. We find a variety of quantum states of light that can be generated by a judicious choice of the input light and electron states. For example, we show how shaping an electron into an energy comb can provide an implementation of a photon displacement operation, allowing, for instance, the generation of displaced Fock and displaced squeezed states. We also show how one can generate a desired Fock state by repeated interactions of electrons with a cavity, followed by measurements. We develop the underlying theory of the interaction of both a single and many consecutive electrons with a common cavity mode. Looking forward, by exploiting the degrees of freedom of arbitrary electron-photon quantum states, we may achieve complete control over the statistics and correlations of output photonic states, leading to the generation of novel quantum states of light.**

**Introduction**

The ability to design and control arbitrary quantum light sources has been a desirable (albeit hard to achieve) goal for many years, especially at the level of a single photon and few photons. Single-photon sources are a key building block for a variety of technologies, such as quantum computing schemes (1-4), metrology (5), teleportation (6) and secure quantum communication (7-9). Entangled states of light composed of two photons or more (e.g. NOON states), and states of light with non-trivial photon statistics (e.g. squeezed states), are a crucial component in numerous applications, like quantum metrology, quantum sensing, quantum imaging, quantum cryptography and continuous variable quantum computing (10-14). Other nonclassical states of light, such as photon-subtracted and photon-added states with non-Gaussian statistics, have also proven valuable for continuous-variable quantum information (15), quantum key distribution (16) and even quantum metrology (17).

The realization of these non-trivial quantum states of light is inherently challenging, increasingly so the more photons involved. Photon entanglement can be realized with the use of nonlinearities, as in spontaneous parametric down conversion (SPDC) (18-19). However, this technique is limited to specific wavelengths, and even for those specific wavelengths, the inefficiency of the nonlinearity process leads to a small throughput of entangled pairs. Other approaches for shaping the quantum state of light have been explored over the years, for example by using atoms launched through a photonic cavity (20-21). Nevertheless, generated photonic states in existing approaches are limited to a relatively small number of photons. We lack methods to create many-photon states of non-trivial statistics. For example, the cat-state with the most photons generated so far had $n = 10$ photons (22), Fock states of only up to $n = 3$ photons were created (23), and similar low photon numbers limit other novel quantum states of light. High photon number states are of great importance for applications such as cluster state quantum computations (24) as well as super-resolved phase sensitivity in quantum metrology (25).

New techniques and methods for overcoming these technological difficulties are much needed. Developing new sources of quantum light states, for various wavelengths, with high throughput and fidelity, could prove to be very useful in the field of quantum optics – allowing new fascinating implementations of quantum technologies. In this work, we propose a scheme for the generation of light with novel quantum states, by utilizing efficient interactions of free-electrons with photons in cavities. Modern developments in electron microscopy tie in very neatly with this, and provide the necessary platform for manipulating light, through its interaction with free electrons, in an effect called photon-induced nearfield electron microscopy (PINEM) (26).

The experimental demonstration of PINEM in the ultrafast transmission electron microscope (UEM) (27-28) motivated the development of the "conventional" PINEM theory (29-31). This theory showed that the entire electron–light interaction can be successfully described (analytically), in a semi-classical manner (29-33), in which the electron is treated quantum mechanically, while the light is treated as a classical field (a strong coherent state). In this scenario, a passing relativistic free electron interacts with an optical field (of some central frequency $\omega$) populated with many photons. The electron energy distribution after the interaction has sharp peaks around its initial energy, with shifts of integer multiples of $\hbar\omega$, thus indicating multiple absorptions and stimulated emissions of the optical field. Such an energy distribution suggests the presence of a quantized ladder of energy levels that the electron moves through as it undergoes the interaction.

This PINEM theory has managed to explain a plethora of new phenomena such as Rabi oscillations of free electrons (33), Ramsey interferometry (34), coherent control (35), stimulated light emission by pre-bunched electrons (36-38), attosecond electron pulse generation (39-41), and electron vortex beam generation (42-43). The PINEM interaction (44) has also enabled new capabilities in electron microscopy, such as improved electron energy loss spectroscopy (45) and various imaging capabilities (46-48,32).

So far, PINEM experiments and theory have all assumed very weak coupling between the electron and the light field, which was assumed to be a strong classical field. However, recent works (49-51) have lifted both assumptions by introducing a new theory of quantum PINEM (QPINEM) (50), in which the light field is quantized as well. This new generalized theory opens the possibility to consider how the electron spectra would behave after interacting with non-classical light sources, as well as investigating such systems when approaching the strong coupling regime. The most recent experimental papers in the PINEM field have gone beyond the weak coupling regime (52-54) and are expected to soon reach the regime needed to observe QPINEM.

Here we propose to use free electrons to create desirable quantum photonic states. We show that by using free electrons, one can generate photon-added states, Fock states, thermal states, displaced coherent states and displaced Fock states. The overarching goal is to eventually design a general scheme to alter the photonic quantum state in controllable ways. For this purpose, we develop the necessary formalism to use the QPINEM interaction for controlling quantum light states. We extend the QPINEM theory with density matrix formalism, and present a robust scheme to handle multiple consecutive interactions of electrons with a common cavity mode. To precisely quantify the electron–photonic-cavity interaction in an arbitrary electromagnetic environment, we develop the macroscopic quantum electrodynamic (MQED) (55-56) framework for the QPINEM interaction with a single photonic mode.

# I. Basic Theory and Results

A general experimental scheme for measuring electron–multi-photon interactions is shown in figure 1a. It demonstrates how the interaction of the free electron and the cavity multi-photon mode (denoted respectively by the quantum states $|\psi^{(i)}\rangle_e$ and $|\psi^{(i)}\rangle_p$) generally result in an entangled quantum state. In the inset of the figure, we mention several photonic structures that would be suitable for efficient interactions (with low losses), thus allowing us to observe the quantum effects discussed here.

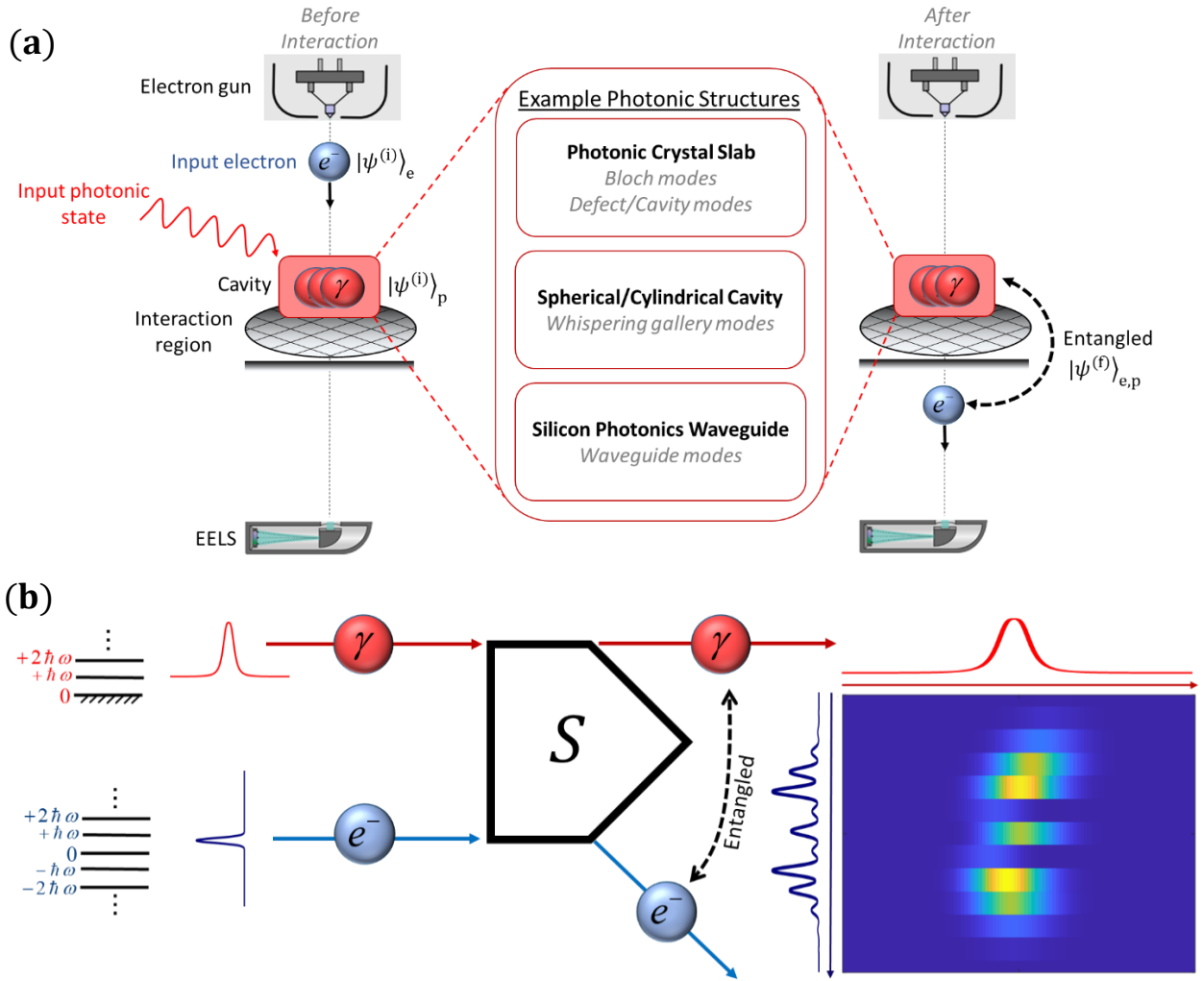

**FIGURE 1: Shaping photonic states of novel quantum statistics using QPINEM interactions.** (a) *Schematic for a physical realization of a QPINEM interaction.* Input electron and photonic states, with specific energy distributions, are generated by the electron gun and some external light source (e.g. a laser pulse) respectively. The interaction takes place between the cavity mode's field

and the electron pulse, where the output states are now, usually, entangled. Finally, the electron is measured by an electron energy loss spectrometer (EELS). Inset lists several optional photonic structures suitable for high-g QPINEM interactions, as in (53-54,57). (**b**) **Interaction scheme of a single QPINEM interaction.** Photonic states can be described by a half-infinite energy ladder (a quantum harmonic oscillator), with respect to a single frequency $\omega$. Electron states can be described by an infinite energy ladder with the same energy steps. Example input and output states are drawn with a photon-electron state probability map at the output.

We start by introducing the state basis. As shown in figure 1b, the electron and photon states are represented by an infinite and half-infinite energy ladders, respectively. More explicitly, the basis of the photonic state is comprised of the Fock states $|n\rangle_p$ (the $n$th step in the photonic energy ladder). The basis of the electron state is comprised of $|k\rangle_e$, the state of an electron with

$$|\psi^{(i)}\rangle = \sum_{\substack{k=-\infty \\ n=0}}^{\infty} c_{k,n}^{(i)} |k, n\rangle. \tag{1}$$

energy $E_0 + k\hbar\omega$ (i.e. the $k$th step in the electron energy ladder), where $E_0$ is the electron's baseline energy. We define the combined electron–photon basis states as $|k, n\rangle = |k\rangle_e \otimes |n\rangle_p$. Generally, as commonly done in all prior works on QPINEM (50-51), we may express any pure input system state as

This description lets us find many properties of the output state, including the exact amplitude coefficients after a QPINEM interaction with a single electron (presented below in equation 11). The pure state description is, however, insufficient when dealing with long chains of QPINEM interactions, i.e. when multiple electrons interact with the same cavity mode, as analyzed in this work. To describe such interactions, we utilize the density-matrix representation for the input state

$$\rho^{(i)} = \sum_{\substack{k,k'=-\infty \\ n,n'=0}}^{\infty} \rho_{k,n,k',n'} |k,n\rangle\langle k',n'|, \tag{2}$$

allowing us to examine not only pure states, as in equation 1, but mixed states as well. Furthermore, density matrix formalism is useful for quantifying and introducing entanglement measures for the states that result from the QPINEM interaction. In this representation, if the electrons interact in a "one-at-a-time" fashion, we can describe the interaction effect of each electron on the cavity using the reduced density matrix of the photons' state that resulted from the preceding electrons. This reduced density matrix is obtained using a partial trace-out of the electron degrees of freedom (if the electron is not measured) or by a projection to a specific electron sub-space (if the electron is measured).

Next, we introduce the Hamiltonian that defines the state basis above by taking the same approximations as in all PINEM theory (29-30) and experiments: (**1**) the magnetic vector potential $A$ is weak relatively to the electron momentum ($|eA| \ll E_0$), allowing us to neglect the diamagnetic ($A^2$) term; (**2**) the electron travels through a charge-free region, so we may take the generalized Coulomb gauge and also assume zero scalar potential; and (**3**) the electron is paraxial, having the majority of its momentum in the $\hat{z}$ direction where its dispersion can be approximated as linear (i.e. constant velocity $v$) (58). These result in the following Hamiltonian

$$\mathcal{H} = -i\hbar v \partial_z + \hbar \omega a^\dagger a + ev\big(A_z(z)a + A_z^\dagger(z)a^\dagger\big), \tag{3}$$

where $a, a^\dagger$ are the photon annihilation and creation operators, which satisfy $[a, a^\dagger] = 1$. The vector potential of a single photon is denoted by $A_z(z) = \sqrt{\frac{\hbar}{2\epsilon_0 \omega}} F_z(z)$, and normalized such that

$$\frac{1}{2\omega} \int d^3 r F^* \frac{d(\omega^2 \varepsilon(r, \omega))}{d\omega} F = 1. \tag{4}$$

This can be directly generalized to an arbitrary optical medium (possibly lossy) using the electromagnetic Green function (59). Recapping equation 3, the first term in the Hamiltonian represents the electron energy under the paraxial approximation, the second term represents the electromagnetic energy for a single light mode, and the third term represents the interaction Hamiltonian.

The system's evolution in time is given by the time evolution operator $U(t)$. In the limit of $t \to \infty$), we get (up to some global phase $e^{i\chi}$ (51) that does not affect observables)

$$S \triangleq U(t \to \infty) = \exp\left[g_{\text{Qu}} e^{-i\frac{\omega}{v}z} a^\dagger - g_{\text{Qu}}^* e^{i\frac{\omega}{v}z} a\right], \tag{5}$$

where the operators $e^{-i\frac{\omega}{v}z}, e^{i\frac{\omega}{v}z}$ can be thought of as the electron energy ladder operators, which we denote as $b, b^\dagger$ respectively. These operators satisfy $b|k\rangle_e = |k-1\rangle_e$ and $b^\dagger|k\rangle_e = |k+1\rangle_e$. Additionally, we define $g_{\text{Qu}}$ with the electric field $E_z$, remembering that in the absence of a scalar potential $E = -\partial A/\partial t$:

$$g_{\text{Qu}} = \frac{e}{\hbar\omega} \int_{-\infty}^{\infty} dz'\, e^{-i\frac{\omega}{v}z'} E_z(z'). \tag{6}$$

The scattering operator $S$ in equation 5 can be written in the form of the displacement operator as $D(bg_{\text{Qu}})$, where $D(\alpha) = \exp(\alpha a^\dagger - \alpha^* a)$. $S$ is a special kind of a displacement operator, however, because its argument $bg_{\text{Qu}}$ is an operator itself. We can expand $D$ and get the matrix elements of $S$:

$$\langle k, n|S|k', n'\rangle = s_{n,n'} \cdot \delta_{k+n,k'+n'}, \tag{7a}$$

$$s_{n,n'} = e^{-\frac{1}{2}|g_{Qu}|^2} g_{Qu}{}^{n-n'} \sqrt{n!\, n'!} \sum_{r=\max\{0,n'-n\}}^{n'} \frac{\left(-|g_{Qu}|^2\right)^r}{r!\,(n'-r)!\,(r+n-n')!}. \tag{7b}$$

The scattering operator $S$ is useful for calculating the density-matrix of the combined output state of the electron and the photon after the interaction

$$\rho^{(f)} = S\rho^{(i)}S^\dagger. \tag{8}$$

Recall we wish to study how the electron state can be exploited to control the photonic state. For this reason, we want to generally examine systems in which a cavity holds a photonic state that is built gradually, through consecutive interactions with electron pulses. We can formulate this by a recursive procedure. In the case that the output electron is not measured, we trace out the electron's degrees of freedom and obtain

$$\rho_p^{(f,m)} = \text{Tr}_e\left(\rho^{(f,m)}\right) = \sum_{j=-\infty}^{\infty} \langle j|_e \rho^{(f,m)} |j\rangle_e. \tag{9}$$

The index $m = 1,2,3,...$ is used to number the consecutive interactions with individual electrons. If we post-select a specific set of electron energies, then the sum in equation 9 will only be on those electron energies (and their corresponding $|j\rangle_e$ states). Now consider the light field in the cavity interacting with another electron. To introduce our new electron state $|\psi^{m+1}\rangle_e$ we write a new photon-electron density matrix

$$\rho^{(i,m+1)} = |\psi^{m+1}\rangle_e \langle\psi^{m+1}|_e \otimes \rho_p^{(f,m)}. \tag{10}$$

With this new density matrix, we have returned, effectively, to exactly where we were in equation 2 – having a density matrix of the whole photon-electron system. Thus, we can repeat equation 8 to find the resulting density-matrix after the next interaction. We can then trace out the new electron using equation 9, and so on, as many times as the number of electrons interacting with the cavity. It is important to note that this scheme does not yet consider cavity losses - we address this point in the final section. In the presence of losses, we generally expect to arrive at a steady state.

The trace in equation 9 is a critical step in the process, with great consequences on the physics, as it may change the photonic state from pure to mixed, which can be seen as decoherence. Generally, the way we deal with the electron degrees of freedom in this step has a major influence on the photonic state (e.g. electron measurement, post selection or trace out). As we will demonstrate in the following sections, the effect of this step on the photonic state also greatly depends on the incoming electron state.

Lastly, for the simple case of a single QPINEM interaction, we can find the exact amplitude coefficients of the output state (given a pure input state, as in equation 1) to be:

$$c_{k,n}^{(f)} = \sum_{n'=0}^{\infty} c_{k+n-n',n'}^{(i)} s_{n,n'} \qquad (11)$$

where $c_{k,n}^{(f)}$ is the amplitude coefficient of the ket state $|k,n\rangle$ in the output state, and where recall $s_{n,n'}$ is defined in equation 7b.

## II. Electron Interaction with a Coherent State: Generalizing the Conventional PINEM Interaction

Perhaps the simplest demonstration of the implications of the quantum PINEM interaction for creating novel photonic states is seen using the conditions that are already prevalent in conventional PINEM experiments. Our input photonic state is a coherent state $|\alpha\rangle_{\text{p}}$, our input electron state is an electron with baseline energy $E_0$ ($|0\rangle_{\text{e}}$) which we denote from here on as a "*delta*" electron, $|\delta\rangle_{\text{e}}$ (a single peak - "delta" - in the $k$-state space).

Usually, we would have a very strong coherent state in the photonic cavity ($|\alpha|^2 \gg 1$), with very weak coupling during the interaction ($|g_{\text{Qu}}|^2 \ll 1$). Under such assumptions, the interaction results in separable output states, where the photonic state remains approximately $|\alpha\rangle_{\text{p}}$, and the separable electron state gives the known Bessel probabilities $J_k^2(2|g|)$ (30-31), where $g$ is the conventionally defined interaction strength, and relates to $g_{\text{Qu}}$ by $g = g_{\text{Qu}}|\alpha|$.

An intriguing phenomenon occurs when looking at strong coupling, or equivalently, at very weak photonic coherent states. Both cases result in a substantial change to the photonic distribution. Using equation 11, and plugging in the appropriate input amplitude coefficients for our given setup, we get that the output amplitude coefficients are:

$$c_{k,n}^{(\text{f})} = \begin{cases} e^{-\frac{|g_{\text{Qu}}|^2 + |\alpha|^2}{2}} \alpha^{k+n} g_{\text{Qu}}^{-k} \sqrt{n!} \sum_{r=\max\{0,k\}}^{k+n} \frac{\left(-|g_{\text{Qu}}|^2\right)^r}{r!(k+n-r)!(r-k)!} & k+n \geq 0 \\ 0 & k+n < 0 \end{cases} \quad (12)$$

Perhaps a more informative expression can be found by writing the photonic state after a post-selection of an electron in energy bin $k$

$$|\psi^{(\text{f})}\rangle_{\text{p}} \cong \sum_{r=\max\{-k,0\}}^{\infty} \frac{\left(-\alpha|g_{\text{Qu}}|^2\right)^r}{r!(r+k)!} \left[(a^\dagger)^r |\alpha\rangle_{\text{p}}\right]. \quad (13)$$

(Since this is the result of post-selection, $|\psi^{(f)}\rangle_p$ has to be explicitly normalized). What we get, in fact, is an infinite sum of "photon-added" coherent states (60). Experimentally, one could realize this scheme by post-selecting the light according to a measurement of the electron energy, at a certain energy bin $k$. This way the electron is also used for heralding. Such electron energy measurements are regularly performed in electron energy loss spectroscopy (EELS) in electron microscopy, however they are only rarely synchronized for coincidence measurements (61).

Figure 2a shows the interaction scheme for the considered setup. In figure 2b, we present the output photon–electron probability map, along with some visualization of the post-selection process (effectively taking a "slice" out of the map). In 2c, we present the photonic states resulting from the highlighted post-selections. In figures 2d and 2e, we present the same interaction results, but for a higher coupling strength.

This test case is rather exciting, because it provides a simple and familiar setup (PINEM-like) that results in non-Gaussian photon statistics. Additionally, this is an interesting method of generating coherent states shifted up by some number of photons (though, we get a sum of these shifted states, and not a single one). In figure 2c, for a weak interaction strength, we show two such examples of shifted coherent states. By tweaking $g_{Qu}$ and the post-selected $k$, we can shift the mean number of photons by different values. Additionally, in figure 2e, we show that for higher interaction strengths we get more unusual and rich photonic distributions (e.g. super-poissonian), arising from the superposition in the sum of equation 13.

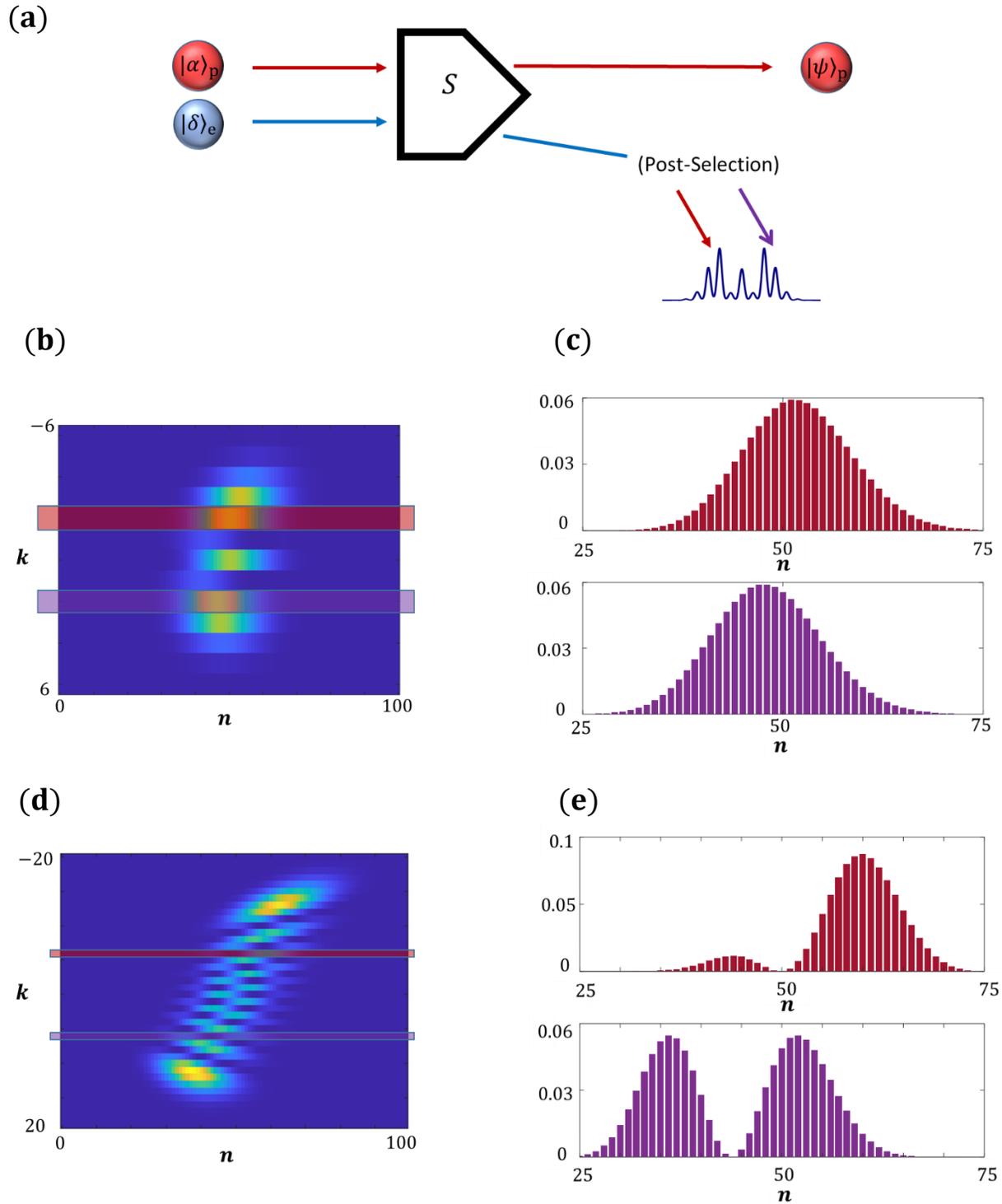

**FIGURE 2: Electron Interaction with a Coherent State: controlling the photonic state by post-selection of the free-electron.** (**a**) **Interaction scheme.** Like in the conventional PINEM case, our input photonic state is a coherent state $|\alpha\rangle_p$, our input electron state is a baseline energy electron, denoted $|\delta\rangle_e \triangleq |0\rangle_e$. At the output of the system, we post-select a specific electron

energy. (**b**) **Output photon-electron probability map.** Showing the entanglement between the two sub-systems. Slices (red and purple) visualize the act of post-selection, as shown in the interaction scheme. (**c**) **Resulting photonic states after post-selection.** We get Poissonian-looking probabilities, around varying mean photon numbers. Bar colors match the previous plot. Red is for post-selected $k = -2$ and purple is for $k = 2$. The initial photonic state used was $|\alpha^2 = 50\rangle_p$ and the interaction strength $g_{Qu} = 0.25i$. (**d**, **e**) The same as (b, c), but for an interaction strength of $g_{Qu} = 1i$ and post-selected energies $k = \pm 6$ (like before, red denotes electron energy loss, and purple denotes gain). We get much richer entanglement for the stronger interaction, as well as more complex photonic states after post-selection.

## III. Creation of Photonic Fock States

We now present a method to create photonic Fock states using consecutive photon-electron interactions and projections, through the electron measurement. This time, we look at a sequence of QPINEM interactions, thus chaining the simple block model previously presented in figure 1b. It is important to note that this time we do not post-select the electron energy, but rather directly measure it. Meaning, we start with an empty cavity, let an electron interact with it, measure the electron's energy, let another electron interact with the cavity and so forth. Our setup includes an initial photonic state of an empty cavity, $|0\rangle_p$, and a "*delta*" electron, $|\delta\rangle_e$. This gives us the very simple initial state $|\psi^{(i)}\rangle = |\delta\rangle_e \otimes |0\rangle_p$, which is in fact a single basis state $|k = 0, n = 0\rangle$.

Examining the explicit expansion of $S$ (as in equation 7a), one finds that it conserves $k + n$ before and after the interaction, per basis state. This means, that since $k + n = 0$ before the interaction, and our input state is a single basis state, then $k + n$ will remain zero after the interaction as well. Once we project on an electron energy bin $k'$, we can uniquely determine that the photonic state can only be $|n' = -k'\rangle_p$ - a pure Fock state. One can generalize the above idea to any general photonic Fock state, and this logic will still hold. Meaning, if we start with a photonic Fock state (an empty cavity, for example), the state will remain a Fock state so long as we follow the scheme described.

It is interesting to note that for an empty cavity, we can only gain photons (or have no change in the photon number). But, if we start with some non-zero Fock state, and measure the electron, we may lose photons too. However, due to the inherent asymmetry of the interaction (50), on average, the photonic state will always gain energy. Which means that given enough interactions, we will reach our desired photon number with probability 1, and stop. While it is guaranteed we will arrive to our desired state, it is not guaranteed when. The arrival time is stochastic, just as the electron measurement is. However, we can calculate the expected number of interactions needed. For the case of *"delta"* electrons, the mean photonic energy gain per interaction is $|g_{\text{Qu}}|^2$ (in units of $\hbar\omega$, or $n$). So, if we started with an empty cavity, and wanted to build a photonic Fock state $N_{\text{goal}}$, we would need, on average, $N_{\text{goal}}/|g_{\text{Qu}}|^2$ interactions to do so. This requires a sufficiently high $Q$-factor cavity, which we discuss further in the final section.

This stochastic process is shown schematically in figure 3a below, where we show an example of the measured electron energies, along with the new photonic Fock state it creates. In figure 3b we show two simulations of the process. We note that if we were to pick a smaller $g_{\text{Qu}}$, then each iteration would yield a smaller energy change, but would allow us to get a finer control over the photonic state, and not get the overshoot we see in the simulations below.

The motivation to such a shaping scheme is the creation of very large number Fock states. These states enable many applications in quantum information, as they can be used to create displaced Fock states (see below in section VI), enable non-Gaussian photon statistics, have many uses in quantum spectroscopy (62), and form a basis for any quantum state.

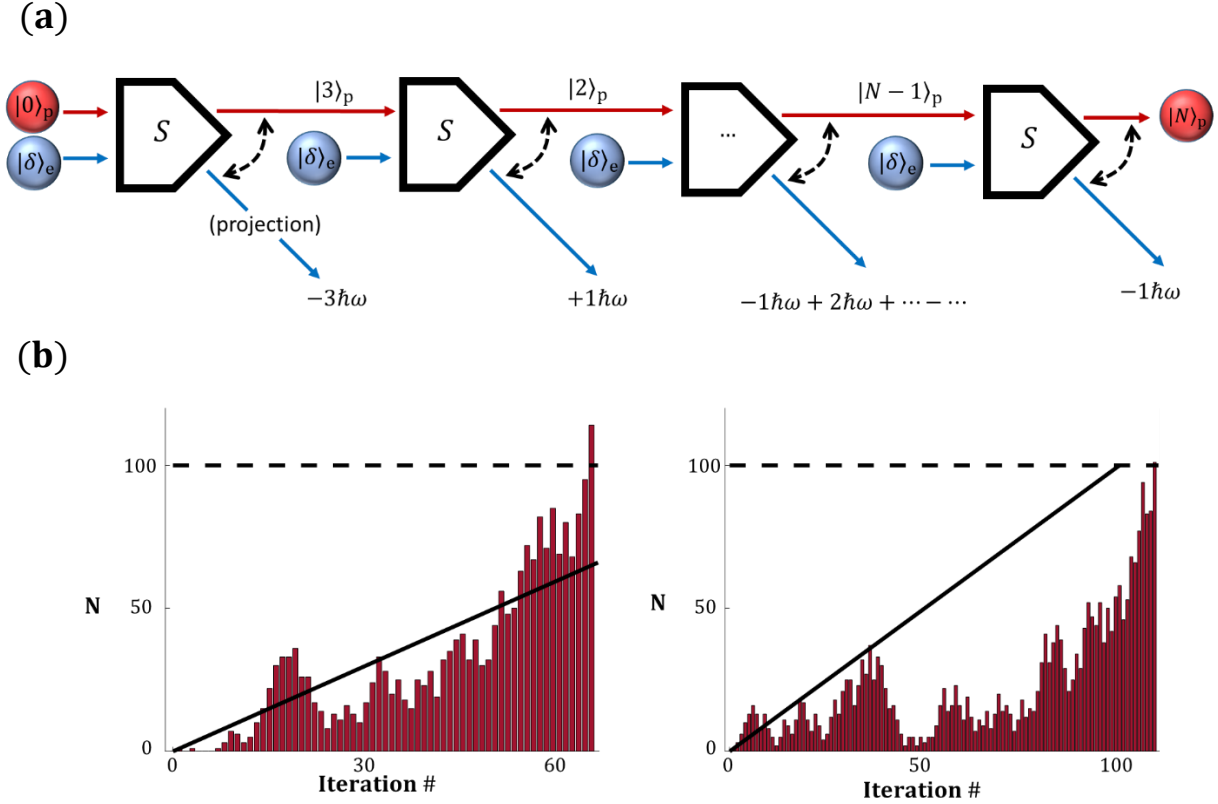

**FIG 3: Creation of a photonic Fock state.** Iterative projection of QPINEM interactions until the desired Fock state is achieved. (**a**) **Interaction scheme.** The input photonic state is that of an empty cavity, the vacuum state. The input electron state is a "*delta*". After each interaction, we measure the electron energy gain / loss, which equals the photonic energy lost or gained respectively. Thus, a desired Fock state may be achieved, after enough iterations. (**b**) **Two examples of the shaping process.** The black dashed line represents the goal Fock state. The red bars show the current photon number, per interaction, and the solid black line – the "mean process", if the photon could go up exactly $|g_{Qu}|^2$ energy steps every interaction. The left process reaches the goal state faster than the average expected growth, and the right reaches it more slowly. Both plots simulate $g_{Qu} = 1i$ and a goal Fock state $|N_{goal} = 100\rangle$. For these values, we expect an average of 100 steps to achieve the goal state.

### IV. Thermalization of Coherent Photonic States

There are already many classical examples of generating thermal states of light, such as using a rotating diffuser (63), and such as how in free-electron lasers, spontaneous emission from many free electrons can become thermal (64). Here, within our framework, we consider a setup identical to section II (a coherent photonic state and a "*delta*" electron state). But, instead of post-selecting the electron energy, we trace out its degrees of freedom by not measuring it, and we repeat this process many times. This leads to the decoherence of the light field, which we show here to result in the thermalization of the photon statistics after many interactions. Consequently, an effective temperature emerges in the cavity. The interaction scheme is presented in figure 4a. We show the state evolution of this process in figure 4b in both linear and log scales. Since a real thermal state $|\theta\rangle_\text{p}$ has an exponential distribution $p_n = (1 - e^{-\theta})e^{-n\theta}$, we expect it to be linear in log scale, which we indeed see in 4b. Figure 4c is a scatter plot that presents the Mandel Q parameter and $\theta = \hbar\omega/k_B T$ versus the average number of photons – showing the gradual convergence of these parameters towards those of a true thermal state. Lastly, we generally expect that any initial photonic state will eventually converge into a thermal state, given a large enough number of interactions, and we use coherent states here as a demonstration of this effect.

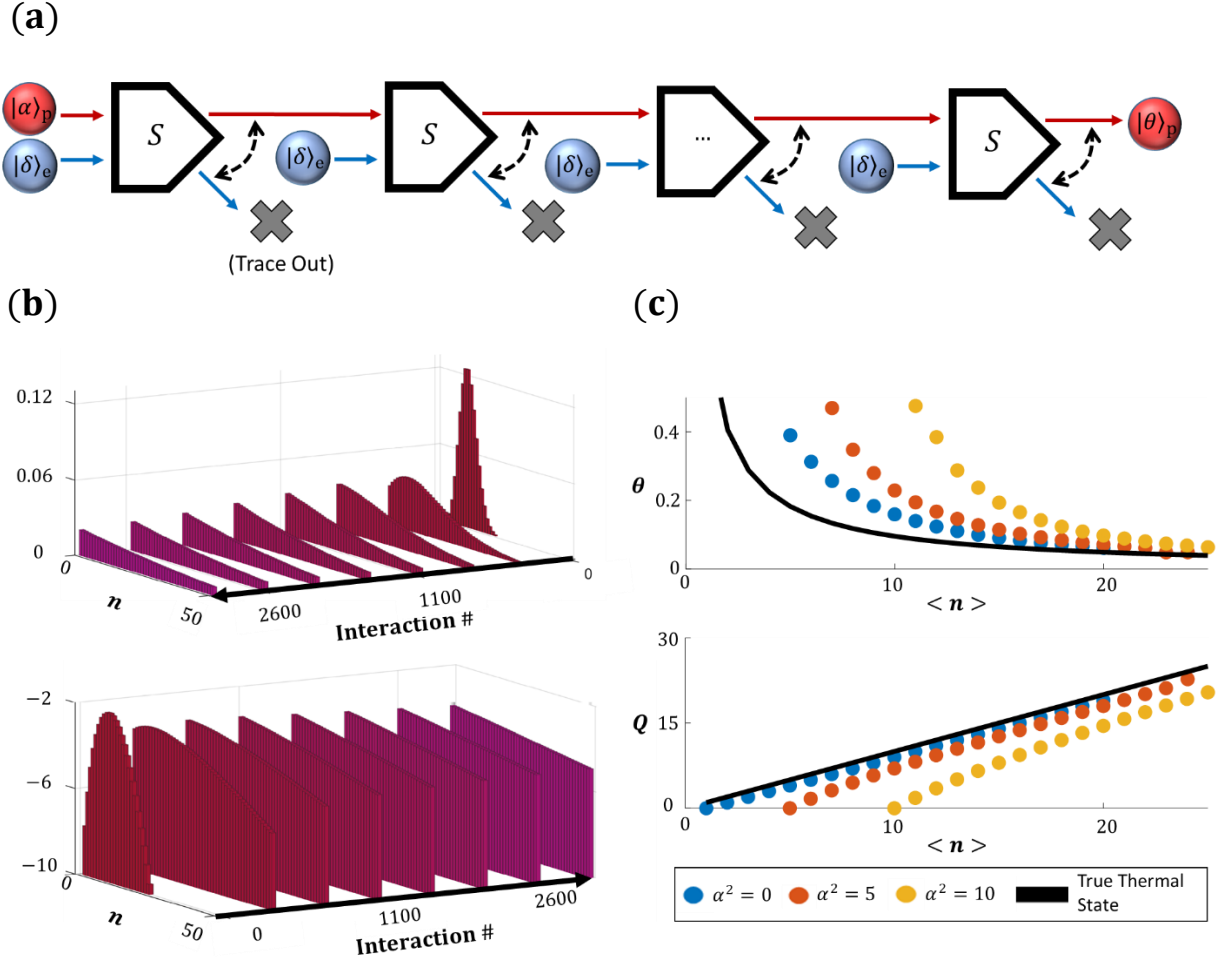

**FIGURE 4: Thermalization of Photonic Coherent States. (a) Interaction scheme.** Input coherent photonic state and a "delta" electron. After each interaction, we trace out the electron state. **(b) Photonic state evolution.** Top: linear scale. Bottom: Log-scale. Both plots depict an initial photonic coherent state $|\alpha^2 = 10\rangle$ and a coupling strength $g_{Qu} = 0.1i$. After many interactions, the statistics converge to thermal (easily seen by the linearity under log-scale). **(c) Photonic state properties.** Top: The effective $\theta = \hbar\omega/k_B T$ of the photonic state (computed as minus the mean slope in log-scale), for different initial coherent states. For the early interactions, this effective slope has no special meaning, as the state has not converged to a thermal state yet. However, it is plotted nevertheless for completeness. Bottom: Mandel Q parameter (65) of the evolving states, for the same three different initial coherent states. Both of these properties visibly converge to that of a true thermal state with the same $\langle n \rangle$. Both plots simulate an interaction strength of $g_{Qu} = 1i$.

## V. Displacement of Photonic Coherent States

We have seen between sections III and IV the fundamental change that the measurement of the electron can have on the final photonic state. This is because an empty cavity is a valid initial state for both setups, yet the resulting distributions are very different (Fock vs thermal). This time, we again compare to section IV, but instead of changing the nature of the measurement, we will change the input electron state. This interaction scheme is presented in figure 5a.

The new electron state we use is an eigenstate of the electron ladder lowering operator $b$. That is, it is some electron state $|c\rangle_e$ for which $b|c\rangle_e = \beta|c\rangle_e$, where $\beta$ is the appropriate eigenvalue. The notation $|c\rangle_e$ stands for "*comb*", as one could imagine such an eigenstate as the limit of an infinite, equally distributed comb of electron energies (in the $k$-state space), with a phase difference of $\beta$ between each two consecutive $k$-states (hence, we require $|\beta| = 1$):

$$|c\rangle_e = \lim_{K,K' \to \infty} \frac{1}{\sqrt{K + K' + 1}} \sum_{k=-K}^{K'} \beta^k |k\rangle_e. \tag{14}$$

One can prove that under this setup, when applied to our system state, the scattering operator is equivalent to

$$S = D(bg_{\text{Qu}}) \Leftrightarrow D(\beta g_{\text{Qu}}), \tag{15}$$

where now the argument to the displacement operator is a scalar. This makes it very easy to prove that applying $S$ to a coherent photonic state is equivalent to applying an additional displacement, that is

$$S[|c\rangle_e \otimes |\alpha\rangle_p] = |c\rangle_e \otimes [D(\beta g_{\text{Qu}})|\alpha\rangle_p] \cong |c\rangle_e \otimes |\tilde{\alpha} = \alpha + \beta g_{\text{Qu}}\rangle_p, \tag{16}$$

where $|\tilde{\alpha}\rangle_\text{p}$ is a photonic coherent state. Note, however, that experimentally, one cannot generate such an infinite comb (a true eigenstate). Instead, we have simulated a finite electron comb to demonstrate the effect, which still preserves very well the coherence of the state. Also, note that the last equality is true only up to a global phase that does not affect any observables, or the photon statistics.

Figure 5b shows the photonic state evolution. We can visibly notice the up-shifting (in the $n$ ladder) of the photonic state, while not seeing too much degradation of the Poissonian statistics. A wider electron comb (more $|k\rangle_\text{e}$ states) would have resulted in a state even closer to a true coherent one.

In figure 5c we show two properties of the evolving photonic state, as it goes through more and more interactions. On the top, we have the effective $\alpha$ of the photonic state. We see that, as expected, it goes up linearly by $\beta g_\text{Qu}$. Generally, we would need to draw this $\beta g_\text{Qu}$ shift as a walk on the complex plane, but both $\alpha$ and $\beta g_\text{Qu}$ have been chosen to be real and positive, for simplicity. On the bottom of figure 5c, we show the Mandel $Q$ parameter (65) of the photonic distribution, for various electron comb lengths. It is exactly 0 for a perfect Poissonian distribution, and gets higher the wider the variance is, compared to the mean. We see that $Q$ remains very low throughout the whole process, for sufficiently long electron combs, indicating the closeness to the ideal Poissonian statistics.

While we have demonstrated the displacement of coherent states in this section, the result is in fact much more general and useful. Up to equation 15, we have not assumed anything about the photonic state. This means that we can use electron combs as a very useful tool to displace any photonic state, regardless of its distribution. This is precisely what we show in the next section, where we generate the well-known displaced Fock state.

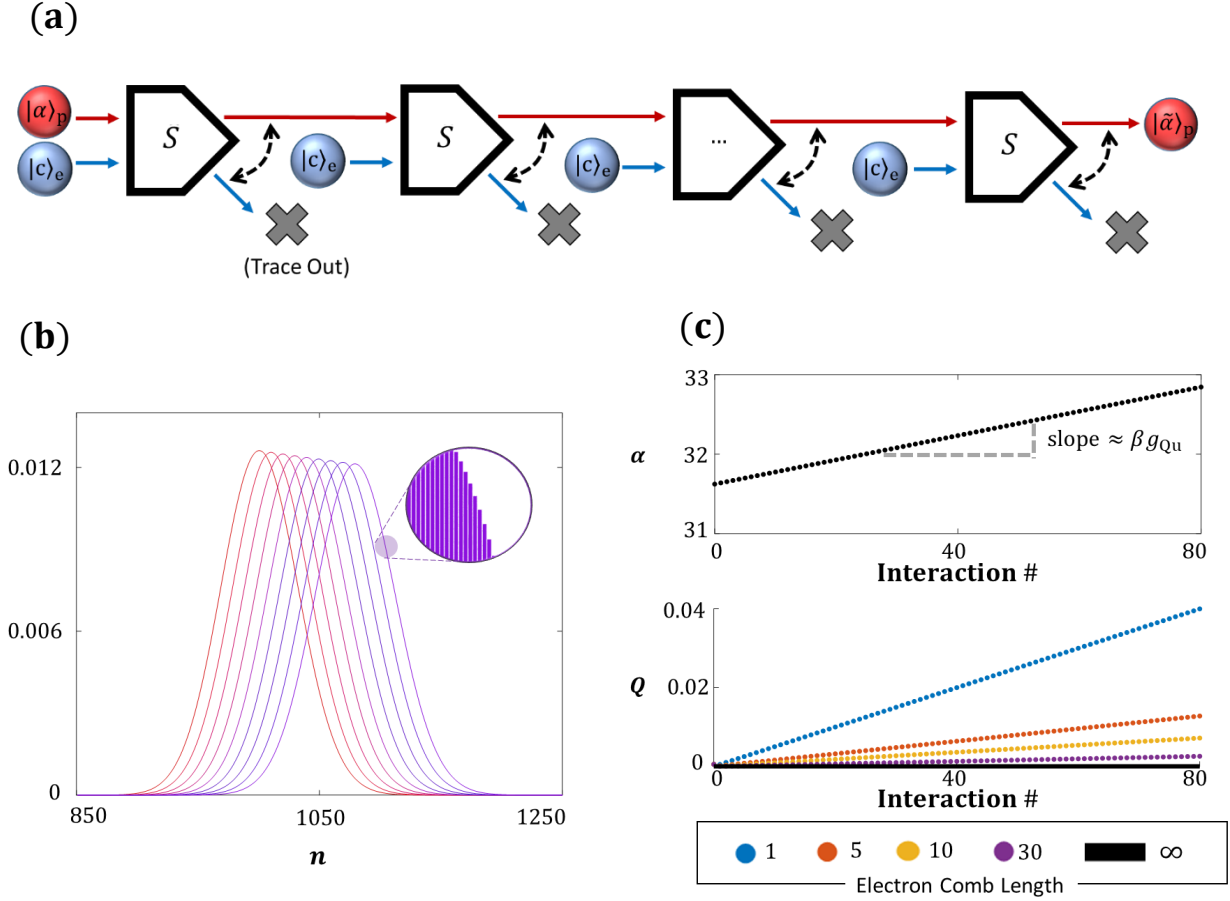

**FIGURE 5: Displacement of Photonic Coherent States.** (**a**) **Interaction scheme.** Input photonic state is a coherent state. Input electron state is a "comb". After each interaction, we trace out the electron and introduce a new electron comb. (**b**) **Photonic state evolution.** Evolves from red to purple, using an electron comb of 30 states. The state maintains quite well its Poissonian distribution while gradually gaining energy from each interaction. (**c**) Top: The effective α evolution (computed as $\sqrt{\langle n \rangle}$) of the photonic state, which evidently grows linearly (with a slope of 0.0153), very close to the theoretical slope of $\beta g_{Qu} = 0.0158$. Bottom: The Mandel Q parameter, for different comb lengths (number of electron states), which remains very low throughout the whole process, for long enough combs. All of the simulation above are for an initial $\alpha = \sqrt{1000}$, a comb with eigenvalue $\beta = -i$ and an interaction strength $g_{Qu} = 0.0158i$.

## VI. Creation of Displaced Fock States

Displaced Fock states (66-68) have proven useful for the direct measurement of Wigner functions (69-72), quantum dense coding (73), and for fundamental tests of quantum mechanics (74). Once the phenomenon in section V is understood, the creation of photonic displaced Fock states is very simple. As we show in figure 6a, the interaction scheme is very similar to that in section V, except now our input photonic state is a Fock state and not a coherent state. Recall, as in equation 15, that a QPINEM interaction with an electron comb (an eigenstate of $b$) is equivalent to applying $D(\beta g_{Qu})$ to the input photonic state. These displacements add up between interactions. That means that after, say, $M$ interactions, our output photonic state will be

$$[D(\beta g_{Qu})]^M |N_i\rangle_p = D(M\beta g_{Qu})|N_i\rangle_p \triangleq |N_i, \alpha = M\beta g_{Qu}\rangle, \tag{17}$$

where we use the standard notation of the form $|N, \alpha\rangle$ for displaced photonic Fock states. The first argument represents the Fock state that we displace, and the second argument tells us by how much. Do note, however, that like in section V, we cannot obtain true displacement, because that would require an infinite electron comb – an true eigenstate of $b$. In our simulation, we again use finite combs and get a very good result, nonetheless.

Another, more explicit expression for the output photonic state may be found from equation 17 as

$$\sum_{r=0}^{N_i} (-\alpha)^{N_i - r} \frac{N_i!}{r!(N_i - r)!} [(a^\dagger)^r |\alpha\rangle_p], \tag{18}$$

where $|\alpha\rangle_p$ is a coherent photonic state defined by $\alpha = M\beta g_{Qu}$, like in equation 17. We again find a sum of photon-added coherent states, like in equation 13 in section II, albeit a finite one, with different summing coefficients.

In figure 6b we show several generated displaced Fock states, and their evolution over multiple interactions with electron combs. For each initial Fock state $N_i$, we end up with $N_i + 1$ peaks that seem roughly Poissonian. The more interactions we perform, the more we displace the Fock state – thus giving it more energy (for our chosen $g_{Qu}$ and $\beta$), and giving each peak a more clear distinctive shape.

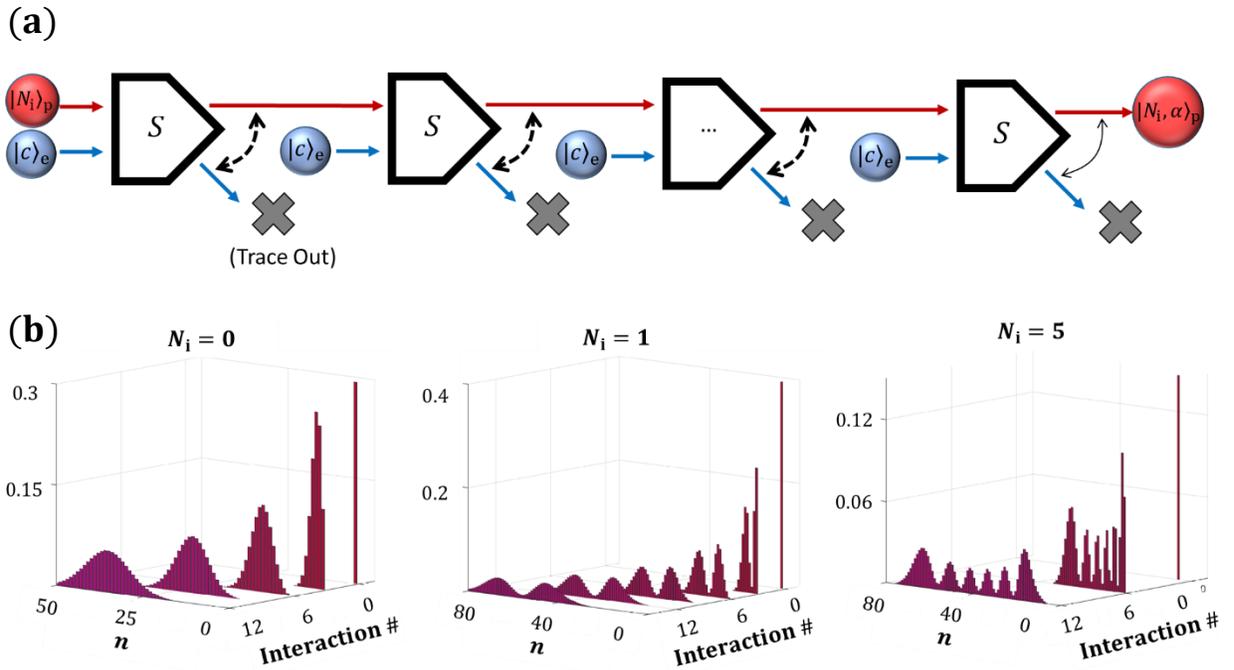

**FIG 6: Creation of displaced Fock states. (a) Interaction scheme.** Input photonic state is a Fock state. Input electron state is a "comb". After each interaction, we trace out the electron and introduce a new electron comb. **(b) Photonic state evolution.** Plotted for various initial photonic Fock states, where for each initial Fock state $N_i$, we end up with $N_i + 1$ roughly Poissonian-looking peaks. In all of the above plots, the interaction strength used is $g_{Qu} = 0.5i$.

## VII. Discussion

We now move to consider more realistic parameters to test the above concepts experimentally. The first important parameter is the cavity lifetime. The existence of a finite cavity lifetime $\tau$ is due to the different leakage channels of the cavity. For the scenarios in which we perform multiple electron interactions, we denote $\Delta t$ as the time between arrivals of consecutive electrons. In this case, we can directly generalize the process expressed in equations 8 - 10 by using a Lindblad master equation, instead of just the Schrödinger equation. Specifically, we need to update the photonic density matrix between consecutive electrons (between equations 9 and 10) as

$$\rho_\text{p} \to \rho_\text{p} - \frac{\Delta t}{\tau}\left(a^\dagger a \rho_\text{p} + \rho_\text{p} a^\dagger a - 2a\rho_\text{p} a^\dagger\right). \tag{19}$$

Note that the results above (figures 3-6) will hold under the condition $\Delta t/\tau \ll 1$ (for a single interaction, or $M\Delta t/\tau \ll 1$ for $M$ consecutive interactions), i.e. a long cavity lifetime, or a short duration between the arrival of consecutive electrons. For example, $\Delta t$ is related to the laser repetition rate in laser-driven electron emission (27-28).

Let us consider typical parameters in current PINEM experiments. Recent PINEM works demonstrated free-electron interactions with cavities having a lifetime of up to 260/340 fs (53-54). Our UTEM setup, and others, currently perform PINEM experiments with electron currents of < 0.1 nA, which corresponds to $\Delta t > 1ns$. These values result in $\Delta t/\tau$ much greater than one, i.e. the photonic state in the cavity will decay before the next electron arrives. One way to reduce $\Delta t/\tau$ is to work at higher currents, which can be done by either increasing the number of electrons per pulse (estimated to reach thousands in femtosecond pulses and millions in nanosecond pulses), or increasing the repetition rates (i.e. approaching GHz instead of the current MHz). The resulting current will then reach ~100 nA, which is typical in certain electron microscopes (especially ones used for cathodoluminescence (75-76)). Such currents correspond to a picosecond lifetime for which $\Delta t/\tau \ll 1$ could hold. Another way to reduce $\Delta t/\tau$ is to use higher Q cavities, which will require working with laser excitations of narrower bandwidths, as in nanosecond lasers (77). Cavities with Q of $10^8$ are used in state-of-the-art experiments with CW lasers (78-79). Of special interest for free-electron experiments are cavities of extended geometries that can enable elongated and prolonged interactions (52,80) to reach a higher $g_{\text{Qu}}$. Such cavities can be designed using phenomena such as photonic bound states in the continuum (81-82), recently reaching Q of $5 \cdot 10^5$ by exploiting topological phenomena (83).

While designated experiments for our purpose of shaping the quantum statistics of light are still beyond reach, there is significant progress in that direction in very recent experiments. What is likely the most promising avenue is combining PINEM with photonic cavities (53-54), and PINEM with elongated structures (52,80).

The same experimental platforms we consider here were proposed and used before to create novel nanophotonic light sources, driven by free-electrons, while so far without controlling the quantum photon statistics. This way, our formalism and predictions can contribute to the growing interest of recent years in novel free-electrons light sources based on nanophotonic structures and high-Q cavities (84). For example, Smith-Purcell sources such as the "light-well" (85), some requiring low electron energies (86), reaching the IR telecom wavelength (87), and the deep UV (88). Other interaction geometries with nanophotonic structures and materials involve metamaterials (89-90), metasurfaces (91), and 2D materials (92-94) for generating light in various spectral regimes up to X-rays and gamma-rays. Such nanophotonic light sources are attractive due to their tunable wavelength. Our work presents the prospect of controlling additional properties of light in future light sources, particularly, their photon statistics. Our analysis points to an additional advantage of free-electron light sources – the electron interaction provides a way for heralding – post-selecting by the electron tells us when the photon state was created. We further note that the formalism we presented is general: it captures many effects beyond what was discussed in this work. For example: Cherenkov radiation from multiple consecutive electrons (starting with a vacuum photonic state).

It is interesting to compare our approach for shaping quantum light with previous ideas that launch beams of atoms (rather than free electrons) through a photonic cavity (20-21). Specifically, shaping quantum light with atom beams would entail the iterative preparation of their states before interaction with the photonic cavity, and the measurement of the atomic state after the interaction; the atoms are modeled by two-level-systems. The main strength of this concept is the ability to plan the exact needed two-level-system states to achieve any desired photonic state (20). However, there are also certain limitations to this concept in comparison with free-electrons: (1) The measurement of the two-level-system involves post-selection at every iteration, so that each atom must come out at the ground state every time, or else the process must be restarted from vacuum. The free-electron interaction schemes presented above do not require starting at the vacuum state. (2) Due to the nature of the two-level-systems, only up to one photon can be gained per interaction, in contrast with free electrons that can exchange a large number of photons every interaction.

Looking at the bigger picture, it is valuable to ask what quantum photonic states could eventually be created from general free-electron–cavity interactions. This is connected to a fundamental question in mathematical physics, because the underlying interaction is of an energy ladder and a harmonic oscillator. A big question is whether such an interaction could provide a platform to shape any desirable photon statistics, or whether it is inherently limited to some subset of states.

Lastly, there is also a fundamental question that arises from our work and connects to the act of measurement in quantum mechanics: is it correct to make a partial trace-out of the electron after each interaction? It may be that in rapid interactions, the first electron is not yet "measured" by the time the second electron interacts with the system. Mathematically, this question amounts to asking whether the formulation we presented in equations 8-10 is still accurate if some of the electrons are only measured after another electron interacts with the cavity (95). Physically, this means that consecutive electrons become entangled (50-51,96), which opens up possibilities to implementing quantum gates (97) for manipulation of the electron quantum state.

**References**

(1) Menicucci, N. C., Van Loock, P., Gu, M., Weedbrook, C., Ralph, T. C., & Nielsen, M. A. Universal quantum computation with continuous-variable cluster states. *Physical Review Letters*, *97*(11), 110501 (2006).

(2) Kok, P. et al. Linear optical quantum computing with photonic qubits. *Reviews of Modern Physics* **79**, 135–174 (2007).

(3) Aspuru-Guzik, A. & Walther, P. Photonic quantum simulators. *Nature Physics* **8**, 285–291 (2012).

(4) Khasminskaya, S., Pyatkov, F., Słowik, K., Ferrari, S., Kahl, O., Kovalyuk, V., ... & Gol'Tsman, G. Fully integrated quantum photonic circuit with an electrically driven light source. *Nature Photonics*, *10*(11), 727-732 (2016).

(5) Cheung, J. et al. The quantum candela: a re-definition of the standard units for optical radiation. *Journal of Modern Optics* **54**, 373–396 (2007).

(6) Nilsson, J., Stevenson, R. M., Chan, K. H. A., Skiba-Szymanska, J., Lucamarini, M., Ward, M. B., ... & Shields, A. J. Quantum teleportation using a light-emitting diode. *Nature Photonics*, *7*(4), 311-315. (2013).

(7) Ekert, A. K. Quantum cryptography based on Bell's theorem. *Physical Review Letters* **67**, 661–663 (1991).

(8) Brendel, J., Gisin, N., Tittel, W., & Zbinden, H. Pulsed energy-time entangled twin-photon source for quantum communication. *Physical Review Letters*, *82*(12), 2594. (1999).

(9) Beveratos, A. et al. Single photon quantum cryptography. *Physical Review Letters* **89**, 187901 (2002).

(10) Lugiato, L. A., Gatti, A., & Brambilla, E., Quantum imaging. *Journal of Optics B: Quantum and Semiclassical Optics* **4** S176 (2002).

(11) Gisin, N., Ribordy, G., Tittel, W., & Zbinden, H., Quantum cryptography. *Reviews of Modern Physics*, **74**(1), 145 (2002).

(12) Braunstein, S. L., & Van Loock, P., Quantum information with continuous variables. *Reviews of Modern Physics*, **77**(2), 513 (2005).

(13) Giovannetti, V., Lloyd, S., & Maccone, L., Advances in quantum metrology. *Nature Photonics*, **5**(4), 222 (2011).

(14) Degen, C. L., Reinhard, F., & Cappellaro, P., Quantum sensing. *Reviews of Modern Physics* **89**(3), 035002 (2017).

(15) Walschaers, M., Fabre, C., Parigi, V., & Treps, N., Entanglement and Wigner function negativity of multimode non-Gaussian states. *Physical Review Letters*, **119**(18), 183601 (2017).

(16) Barnett, S., Quantum information (Vol. 16). *Oxford University Press* (2009).

(17) Orlov, A. A., Krylova, A. K., Zhukovsky, S. V., Babicheva, V. E., & Belov, P. A., Multiperiodicity in plasmonic multilayers: general description and diversity of topologies. *Physical Review A*, **90**(1), 013812 (2014).

(18) Valencia, A., Ceré, A., Shi, X., Molina-Terriza, G., & Torres, J. P. Shaping the waveform of entangled photons. *Physical Review Letters*, **99**(24), 243601 (2007).

(19) Baek, S. Y., Kwon, O., & Kim, Y. H. Temporal shaping of a heralded single-photon wave packet. *Physical Review A*, **77**(1), 013829. (2008).

(20) Vogel, K., Akulin, V. M., & Schleich, W. P. Quantum state engineering of the radiation field. *Physical Review Letters*, **71**(12), 1816. (1993).

(21) Raimond, J. M., Brune, M., & Haroche, S. Manipulating quantum entanglement with atoms and photons in a cavity. *Reviews of Modern Physics*, **73**(3), 565. (2001).

(22) Gao, W. B., Lu, C. Y., Yao, X. C., Xu, P., Gühne, O., Goebel, A., ... & Pan, J. W., Experimental demonstration of a hyper-entangled ten-qubit Schrödinger cat state. *Nature Physics*, **6**(5), 331-335 (2010).

(23) Cooper, M., Wright, L. J., Söller, C., & Smith, B. J., Experimental generation of multi-photon Fock states. *Optics Express*, **21**(5), 5309-5317 (2013).

(24) Tokunaga, Y., Kuwashiro, S., Yamamoto, T., Koashi, M., & Imoto, N., Generation of high-fidelity four-photon cluster state and quantum-domain demonstration of one-way quantum computing. *Physical Review Letters*, **100**(21), 210501 (2008).

(25) Mitchell, M. W., Lundeen, J. S., & Steinberg, A. M., Super-resolving phase measurements with a multiphoton entangled state. *Nature*, **429**(6988), 161-164 (2004).

(26) Barwick, B., Flannigan, D. J., & Zewail, A. H. Photon-induced near-field electron microscopy. *Nature* **462,** 902 (2009).

(27) Lobastov, V. A., Srinivasan, R., & Zewail, A. H. Four-dimensional ultrafast electron microscopy. *Proceedings of the National Academy of Sciences* **102,** 7069-7073 (2005).

(28) Zewail, A. H. Four-dimensional electron microscopy. *Science* **328,** 187-193 (2010).

(29) García de Abajo, F. J., Asenjo-Garcia, A., & Kociak, M. Multiphoton absorption and emission by interaction of swift electrons with evanescent light fields. *Nano Letters* **10,** 1859-1863 (2010).

(30) Park, S. T., Lin, M., & Zewail, A. H. Photon-induced near-field electron microscopy (PINEM): theoretical and experimental. *New Journal of Physics* **12,** 123028 (2010).

(31) De Abajo, F. G., Barwick, B., & Carbone, F. Electron diffraction by plasmon waves. *Physical Review B* **94**, 041404 (2016).

(32) Piazza, L. et al. Simultaneous observation of the quantization and the interference pattern of a plasmonic near-field. *Nature Communications* **6,** 6407 (2015).

(33) Feist, A., et al. Quantum coherent optical phase modulation in an ultrafast transmission electron microscope. *Nature* **521,** 200 (2015).

(34) Echternkamp, K. E., Feist, A., Schäfer, S., & Ropers, C. Ramsey-type phase control of free-electron beams. *Nature Physics* **12,** 1000 (2016).

(35) Vanacore, G. M., et al. Attosecond coherent control of free-electron wave functions using semi-infinite light fields. *Nature Communications* **9,** 2694 (2018).

(36) Gover, A., & Pan, Y. Dimension-dependent stimulated radiative interaction of a single electron quantum wavepacket. *Physics Letters A* **382,** 1550-1555 (2018).

(37) Pan, Y., & Gover, A. Spontaneous and stimulated radiative emission of modulated free-electron quantum wavepackets - semiclassical analysis. *Journal of Physics Communications* **2,** 115026 (2018).

(38) Gover, A., & Yariv, A. Free-Electron–Bound-Electron Resonant Interaction. *Physical Review Letters* **124**, 064801 (2020).

(39) Priebe, K. E., et al. Attosecond electron pulse trains and quantum state reconstruction in ultrafast transmission electron microscopy. *Nature Photonics* **11,** 793 (2017).

(40) Morimoto, Y., & Baum, P. Diffraction and microscopy with attosecond electron pulse trains. *Nature Physics* **14,** 252 (2018).

(41) Kozák, M., Schönenberger, N., & Hommelhoff, P. Ponderomotive generation and detection of attosecond free-electron pulse trains. *Physical Review Letters* **120,** 103203 (2018).

(42) Cai, W., Reinhardt, O., Kaminer, I., & de Abajo, F. J. G. Efficient orbital angular momentum transfer between plasmons and free electrons. *Physical Review B* **98,** 045424 (2018).

(43) Vanacore, G. M., et al. Ultrafast generation and control of an electron vortex beam via chiral plasmonic near fields. *Nature Materials* **18**(6), 573-579 (2019).

(44) Talebi, N. Near-Field-Mediated Photon-Electron Interactions. *Springer*. (2019).

(45) Pomarico, E., et al. meV resolution in laser-assisted energy-filtered transmission electron microscopy. *ACS Photonics* **5,** 759-764 (2017).

(46) Flannigan, D. J., Barwick, B., & Zewail, A. H. Biological imaging with 4D ultrafast electron microscopy. *Proceedings of the National Academy of Sciences* **107,** 9933-9937 (2010).

(47) Yurtsever, A., & Zewail, A. H. Direct visualization of near-fields in nanoplasmonics and nanophotonics. *Nano Letters* **12,** 3334-3338 (2012).

(48) Barwick, B., & Zewail, A. H. Photonics and plasmonics in 4D ultrafast electron microscopy. *ACS Photonics* **2,** 1391-1402 (2015).

(49) Pan, Y., & Gover, A. Spontaneous and stimulated emissions of a preformed quantum free-electron wave function. *Physical Review A* **99**, *052107* (2019).

(50) Kfir, O. Entanglements of electrons and cavity photons in the strong-coupling regime. *Physical Review Letters* **123**, 103602 (2019)

(51) Di Giulio, V., Kociak, M., & de Abajo, F. J. G. Probing quantum optical excitations with fast electrons. *Optica* **6**, 1524 (2019).

(52) Dahan, R., Nehemia, S., Shentcis, M., Reinhardt, O., Adiv, Y., Wang, K., ... & Kaminer, I. Observation of the stimulated quantum Cherenkov effect. *arXiv preprint arXiv:1909.00757* (2019).

(53) Wang, K., Dahan, R., Shentcis, M., Kauffmann, Y., Hayun, A.B., Reinhardt, O., Tsesses, S. and Kaminer, I., Coherent interaction between free electrons and a photonic cavity. *Nature*, **582**(7810), pp.50-54 (2020).

(54) Kfir, O., Lourenço-Martins, H., Storeck, G., Sivis, M., Harvey, T. R., Kippenberg, T. J., ... & Ropers, C. Controlling free electrons with optical whispering-gallery modes. *Nature*, **582**(7810), 46-49 (2020).

(55) Scheel, S., & Buhmann, S. Y. Macroscopic QED-concepts and applications. *arXiv preprint arXiv:0902.3586*. (2009).

(56) Rivera, N., & Kaminer, I., Light-matter interactions with photonic quasiparticles. *arXiv preprint arXiv:2004.07748* (2020).

(57) Sapra, N.V., Yang, K.Y., Vercruysse, D., Leedle, K.J., Black, D.S., England, R.J., Su, L., Trivedi, R., Miao, Y., Solgaard, O. and Byer, R.L., On-chip integrated laser-driven particle accelerator. *Science*, **367**(6473), pp.79-83 (2020).

(58) Additional corrections to the PINEM theory can, in principle, occur due to the change in permittivity along the electron trajectory (due to the fact that **A** and **p** do not commute). However, these corrections can be neglected as long as the electron's $\hat{z}$ momentum is much larger than $\hbar(\partial\varepsilon/\partial z)$.

(59) For a Green function representation, we replace the interaction Hamiltonian with $ev \cdot A$, with $A$ defined as

$$A(r) = \sqrt{\frac{\hbar}{\pi\varepsilon_0}} \int_0^\infty d\omega \frac{\omega}{c^2} \int d^3r' \sqrt{\text{Im}\{\varepsilon(r,r',\omega)\}} [G(r,r',\omega)f(r',\omega) + G^*(r,r',\omega)f^\dagger(r',\omega)].$$

This associates with each $(\boldsymbol{r}, \omega, k)$ a quantum harmonic oscillator, with matching creation $f_k^\dagger(\boldsymbol{r}, \omega)$ and annihilation $f_k(\boldsymbol{r}, \omega)$ operators, satisfying $(f_k(\boldsymbol{r}, \omega), f_{k'}(\boldsymbol{r'}, \omega')) = (f_k(\boldsymbol{r}, \omega), f_{k'}(\boldsymbol{r'}, \omega'))^\dagger = 0$ and $(f_k(\boldsymbol{r}, \omega), f_{k'}^\dagger(\boldsymbol{r'}, \omega')) = \delta_{kk'}\delta(\omega - \omega')\delta(\boldsymbol{r} - \boldsymbol{r'})$. Additional information in (56).

(60) Barnett, S. M., Ferenczi, G., Gilson, C. R., & Speirits, F. C. Statistics of photon-subtracted and photon-added states. *Physical Review A*, **98**(1), 013809 (2018).

(61) Jannis, D., Müller-Caspary, K., Béché, A., Oelsner, A., & Verbeeck, J., Spectroscopic coincidence experiments in transmission electron microscopy. *Applied Physics Letters*, **114**(14), 143101. (2019).

(62) Mukamel, S., Freyberger, M., Schleich, W., Bellini, M., Zavatta, A., Leuchs, G., ... & Barbieri, M., Roadmap on quantum light spectroscopy. *Journal of Physics B: Atomic, Molecular and Optical Physics*, **53**(7), 072002. (2020).

(63) Li, S. W., Li, F., Peng, T., & Agarwal, G. S., Photon statistics of quantum light on scattering from rotating ground glass. *Physical Review A*, **101**(6), 063806. (2020).

(64) Becker, W., & McIver, J. K., Photon statistics of the free-electron-laser startup. *Physical Review A*, **28**(3), 1838. (1983).

(65) The Mandel $Q$ parameter is a parameter in the range of $(-1, \infty)$, that measures the closeness of a given statistic distribution to Poissonian statistics. It is defined as

$$Q = \frac{\langle(\Delta n)^2\rangle - \langle n\rangle}{\langle n\rangle} = \langle n\rangle\big(g^{(2)}(0) - 1\big),$$

where $n$ is the photon number operator, and in the alternative definition, $g^{(2)}$ is the normalized second order Glauber correlation function. $Q$ is zero for a photonic coherent state, is positive for distributions for which the variance is greater than the mean, and negative if it is lesser.

(66) De Oliveira, F. A. M., Kim, M. S., Knight, P. L., & Buek, V., Properties of displaced number states. *Physical Review A*, **41**(5), 2645. (1990).

(67) Lvovsky, A. I., & Babichev, S. A., Synthesis and tomographic characterization of the displaced Fock state of light. *Physical Review A*, **66**(1), 011801. (2002).

(68) Keil, R., Perez-Leija, A., Dreisow, F., Heinrich, M., Moya-Cessa, H., Nolte, S., ... & Szameit, A., Classical analogue of displaced Fock states and quantum correlations in Glauber-Fock photonic lattices. *Physical Review Letters*, **107**(10), 103601. (2011).

(69) Leibfried, D., Meekhof, D. M., King, B. E., Monroe, C. H., Itano, W. M., & Wineland, D. J., Experimental determination of the motional quantum state of a trapped atom. *Physical Review Letters*, **77**(21), 4281. (1996).

(70) Lutterbach, L. G., & Davidovich, L., Method for direct measurement of the Wigner function in cavity QED and ion traps. *Physical Review Letters*, **78**(13), 2547. (1997).

(71) Banaszek, K., Radzewicz, C., Wódkiewicz, K., & Krasiński, J. S., Direct measurement of the Wigner function by photon counting. *Physical Review A*, **60**(1), 674. (1999).

(72) Bertet, P., Auffeves, A., Maioli, P., Osnaghi, S., Meunier, T., Brune, M., ... & Haroche, S., Direct measurement of the Wigner function of a one-photon Fock state in a cavity. *Physical Review Letters*, **89**(20), 200402. (2002).

(73) Podoshvedov, S. A., & Kim, J., Dense coding by means of the displaced photon. *Physical Review A*, **77**(3), 032319. (2008).

(74) Kuzmich, A., Walmsley, I. A., & Mandel, L., Violation of Bell's inequality by a generalized Einstein-Podolsky-Rosen state using homodyne detection. *Physical Review Letters*, **85**(7), 1349. (2000).

(75) Meuret, S., Tizei, L. H. G., Cazimajou, T., Bourrellier, R., Chang, H. C., Treussart, F., & Kociak, M., Photon bunching in cathodoluminescence. *Physical Review Letters*, **114**(19), 197401. (2015).

(76) Polman, A., Kociak, M., & de Abajo, F. J. G. Electron-beam spectroscopy for nanophotonics. *Nature Materials* **18**, 1158–1171 (2019).

(77) Das, P. et al. Stimulated electron energy loss and gain in an electron microscope without a pulsed electron gun. *Ultramicroscopy* **203**, 44 (2019).

(78) Carmon, T. & Vahala, K. J. Modal Spectroscopy of Optoexcited Vibrations of a MicronScale On-Chip Resonator at Greater than 1 GHz Frequency. *Physical Review Letters* **98**, 123901 (2007).

(79) Midolo, L., Schliesser, A. & Fiore, A. Nano-opto-electro-mechanical systems. *Nature Nanotechnology* **13**, 11 (2018).

(80) Adiv, Y. & Kaminer, I. Observation of the Quantum Nature of Laser-Driven Particle Acceleration. In 2020 Conference on Lasers and Electro-Optics (CLEO). IEEE. (2020).

(81) Hsu, C. W., Zhen, B., Lee, J., Chua, S. L., Johnson, S. G., Joannopoulos, J. D., & Soljačić, M. Observation of trapped light within the radiation continuum. *Nature* **499**, 188 (2013).

(82) Hsu, C. W., Zhen, B., Stone, A. D., Joannopoulos, J. D., & Soljačić, M. Bound states in the continuum. *Nature Reviews Materials* **1**, 16048 (2016).

(83) Yin, X., Jin, J., Soljačić, M., Peng, C., & Zhen, B. Observation of topologically enabled unidirectional guided resonances. *Nature*, *580*(7804), 467-471. (2020).

(84) Yang, Y. et al. Maximal spontaneous photon emission and energy loss from free electrons. *Nature Physics* **14**, 894. (2018).

(85) Adamo, G., MacDonald, K. F., Fu, Y. H., Wang, C. M., Tsai, D. P., de Abajo, F. G., & Zheludev, N. I. Light well: a tunable free-electron light source on a chip. *Physical Review Letters* **103**, 113901 (2009).

(86) Massuda, A. et al. Smith–Purcell radiation from low-energy electrons. *ACS Photonics* **5**, 3513 (2018).

(87) Roques-Carmes, C. et al. Towards integrated tunable all-silicon free-electron light sources. *Nature Communications*, **10**, 3176 (2019)

(88) Ye, Y. et al. Deep-ultraviolet Smith–Purcell radiation. *Optica* **6**, 592 (2019).

(89) Adamo, G. et al. Electron-beam-driven collective-mode metamaterial light source. *Physical Review Letters* **109**, 217401. (2012).

(90) Ginis, V., Danckaert, J., Veretennicoff, I., & Tassin, P. Controlling Cherenkov radiation with transformation-optical metamaterials. *Physical Review Letters* **113**, 167402 (2014).

(91) Rosolen, G., Wong, L. J., Rivera, N., Maes, B., Soljačić, M., & Kaminer, I. Metasurface-based multi-harmonic free-electron light source. *Light: Science & Applications* **7**, 1 (2018).

(92) Wong, L. J., Kaminer, I., Ilic, O., Joannopoulos, J. D. & Soljačić, M. Towards graphene plasmon-based free-electron infrared to X-ray sources. *Nature Photonics* **10**, 46 (2016).

(93) Mišković, Z. L., Segui, S., Gervasoni, J. L., & Arista, N. R. Energy losses and transition radiation produced by the interaction of charged particles with a graphene sheet. *Physical Review B* **94**, 125414 (2016).

(94) Rivera, N., Wong, L. J., Joannopoulos, J. D., Soljačić, M., & Kaminer, I., Light emission based on nanophotonic vacuum forces. *Nature Physics*, **15**, 1284–1289 (2019).

(95) We conjecture that our treatment will still be correct in that case, although formally for a period of time, the combined quantum state of the system has a much higher dimensionality.

(96) Mechel, C., Kurman, Y., Karnieli, A., Rivera, N., Arie, A., & Kaminer, I. Imaging the collapse of electron wave-functions: the relation to plasmonic losses. In CLEO: QELS Fundamental Science (pp. FF3M-6). Optical Society of America. (2019).

(97) Reinhardt, O., Mechel, C., Lynch, M., & Kaminer, I. Free electron qubits. In *2019 Conference on Lasers and Electro-Optics (CLEO)* (pp. 1-2). IEEE. (2019).
32

## S1. Assisting Derivations for Section I: Basic Theory and Results

### S1.1. Quantum Electrodynamics of Free Electrons and Light

In this section, we elaborate on the theory and approximations, as well as explicitly derive the QED equations, from section I of the main text (equations 3-6). The Hamiltonian of free-electrons and light is generally given as

$$\mathcal{H} = \mathcal{H}_\text{el} + \mathcal{H}_\text{em} + \mathcal{H}_\text{int},$$

where $\mathcal{H}_\text{el}$ is the electron Hamiltonian, $\mathcal{H}_\text{em}$ is the electromagnetic field Hamiltonian and $\mathcal{H}_\text{int}$ is the Hamiltonian describing the interaction between them. For a system of $N$ relativistic (with negligible spin) electrons, the Hamiltonian is the following Klein-Gordon Hamiltonian

$$\mathcal{H}_\text{el} = \sum_{i=1}^{N} c\sqrt{m^2 c^2 + \mathbf{p}_i^2}.$$

In the presence of an electromagnetic field described by a vector potential $\mathbf{A}(\mathbf{r}_i)$, the Hamiltonian under a minimal coupling transformation contains both the pure electron part and an interaction part. In particular

$$\mathcal{H}_{\text{el}} + \mathcal{H}_{\text{int}} = \sum_{i=1}^{N} c\sqrt{m^2 c^2 + (\mathbf{p}_i + e\mathbf{A}(\mathbf{r}_i))^2}.$$

The vector potential (assuming sufficiently low optical losses) is specified in terms of a complete set of normalized orthonormal modes (labeled by $n$), $\mathbf{F}_n(\mathbf{r})$, and their frequencies $\omega_n$ as

$$\mathbf{A}(\mathbf{r}) = \sum_n \sqrt{\frac{\hbar}{2\epsilon_0 \omega_n}} (\mathbf{F}_n(\mathbf{r}) a_n + \mathbf{F}_n^*(\mathbf{r}) a_n^\dagger),$$

with $\epsilon_0$ being the vacuum permittivity. It is useful to expose the units and clean up this expression by writing it as

$$\mathbf{A}(\mathbf{r}) = \sum_n \mathbf{A}_n(\mathbf{r}) a_n + \mathbf{A}_n^*(\mathbf{r}) a_n^\dagger,$$

where

$$\mathbf{A}_n(\mathbf{r}) = \sqrt{\frac{\hbar}{2\epsilon_0 \omega_n}} \mathbf{F}_n(\mathbf{r})$$

has the dimensions of a vector potential. The electromagnetic Hamiltonian is specified in terms of the modes and their frequencies as

$$\mathcal{H}_{\text{em}} = \sum_n \hbar \omega_n a_n^\dagger a_n.$$

The total Hamiltonian for this problem is already too complicated to solve exactly, due to the appearance of the vector potential in the square root, which makes the free electron term and the interaction term inseparable. We now perform an approximation to separate them, assuming that $|e\mathbf{A}| \ll \sqrt{m^2c^2 + \mathbf{p}^2}$. In that case, we can perform a Taylor expansion of $\mathcal{H}_{el} + \mathcal{H}_{int}$ that results in

$$\mathcal{H}_{el} + \mathcal{H}_{int} \approx \sum_{i=1}^{N} c\sqrt{m^2c^2 + \mathbf{p}_i^2} + \frac{ec}{2\sqrt{m^2c^2 + \mathbf{p}_i^2}}(\mathbf{p}_i \cdot \mathbf{A}(\mathbf{r}_i) + \mathbf{A}(\mathbf{r}_i) \cdot \mathbf{p}_i),$$

where we have ignored the so called *diamagnetic* term proportional to $\mathbf{A}^2$. Additionally, we make another approximation which is commonly made, which is to ignore the non-commutability of $\mathbf{p}$ and $\mathbf{A}$, such that $\mathbf{p}_i \cdot \mathbf{A}(\mathbf{r}_i) = \mathbf{A}(\mathbf{r}_i) \cdot \mathbf{p}_i = -i\hbar\nabla \cdot \mathbf{A}$. This is justifiable when the electron always propagates in regions of constant permittivity. This is because the vector potential and Hamiltonian above are derived under the generalized Coulomb gauge, for which $\nabla \cdot \epsilon\mathbf{A} = 0$. And so, if $\epsilon$ is constant, then the vector potential's divergence is zero. In the case of changing permittivity $\epsilon(z)$, there can be additional corrections, but those can be neglected if the electron's $\hat{z}$ momentum is much larger than $\hbar(\partial\epsilon/\partial z)$.

The Hamiltonian can also be simplified further in the case where we treat the electron under the paraxial approximation, which results from linearizing the dispersion relation of the electron around its central momentum. We therefore restrict the electron Hamiltonian to the space of functions of the form $\psi = e^{i\mathbf{k}_0 \cdot \mathbf{r}}f$, where $f$ is slowly varying ($|\nabla f| \ll |kf|$). In that case, the action of the electron Hamiltonian (taken for a single electron, without loss of generality) on the state is

$$\mathcal{H}_{el}\psi = c\sqrt{m^2c^2 - \hbar^2\nabla^2}e^{i\mathbf{k}_0 \cdot \mathbf{r}}f \approx e^{i\mathbf{k}_0 \cdot \mathbf{r}}c\sqrt{m^2c^2 + \hbar^2\mathbf{k}_0^2 - 2i\hbar^2\mathbf{k}_0 \cdot \nabla f}.$$

Taylor expanding, and noting that our electron is relativistic (so $p/E = v/c^2$) we have

$$e^{i\mathbf{k}_0 \cdot \mathbf{r}}\left(c\sqrt{m^2c^2 + \hbar^2\mathbf{k}_0^2} + \frac{\hbar c\mathbf{k}_0 \cdot (-i\hbar\nabla)}{\sqrt{m^2c^2 + \hbar^2\mathbf{k}_0^2}}\right)f = e^{i\mathbf{k}_0 \cdot \mathbf{r}}\left(c\sqrt{m^2c^2 + \hbar^2\mathbf{k}_0^2} - i\hbar\mathbf{v} \cdot \nabla\right)f$$

$$\equiv e^{i\mathbf{k}_0 \cdot \mathbf{r}}(E_0 - i\hbar\mathbf{v} \cdot \nabla)f.$$

The next step is to approximate $\mathcal{H}_{el}$ as $-i\hbar\mathbf{v} \cdot \nabla$, which is justified since they are the same, up to the identity, in the state space $\psi = e^{i\mathbf{k}_0 \cdot \mathbf{r}}f$. In particular

$$\mathcal{H}_{el}\psi - (-i\hbar\mathbf{v} \cdot \nabla)\psi = e^{i\mathbf{k}_0 \cdot \mathbf{r}}(E_0 - \hbar\mathbf{k}_0 \cdot \mathbf{v})f = (E_0 - \hbar\mathbf{k}_0 \cdot \mathbf{v})\psi,$$

differing only by a constant multiple of the identity, since $\mathbf{v}$ and $\mathbf{k}_0$ are constant for every state. Therefore, our final electron Hamiltonian is

$$\mathcal{H}_{el} \approx \sum_{i=1}^{N}(-i\hbar\mathbf{v} \cdot \nabla_i)$$

Physically, this equation means that the velocity of the electron is $\mathbf{v}$ regardless of its momentum, which is a type of "non-recoil approximation". It is also equivalent to considering the electron as a spinless particle of velocity $v$. In this approximation, we can also simplify the interaction Hamiltonian to be $e\mathbf{v} \cdot \mathbf{A}$. Thus, in the paraxial approximation, the full electron-light Hamiltonian is

$$\mathcal{H} = \sum_{i=1}^{N}(-i\hbar\mathbf{v} \cdot \nabla_i) + \sum_n \hbar\omega_n a_n^\dagger a_n + \sum_{i=1}^{N}\sum_n e\mathbf{v} \cdot \left(\mathbf{A}_n(\mathbf{r})a_n + \mathbf{A}_n^*(\mathbf{r})a_n^\dagger\right).$$

For compactness, we define the following quantity with units of energy

$$W_n = e\mathbf{v} \cdot \mathbf{A}_n(\mathbf{r}),$$

so that the Hamiltonian takes the form

$$\mathcal{H} = \sum_{i=1}^{N}(-i\hbar \mathbf{v} \cdot \nabla_i) + \sum_n \hbar\omega_n a_n^\dagger a_n + \sum_{i=1}^{N}\sum_n (\mathbf{W}_n(\mathbf{r}_i)a_n + \mathbf{W}_n^*(\mathbf{r}_i)a_n^\dagger)$$

Let us now apply the above formalism to a PINEM interaction with a quantized field. We will work under the regular PINEM assumptions. Specifically, we assume that: (1) there is only one electron at a time, and (2) the velocity is along the $\hat{z}$ direction. We further take the whole problem to be one-dimensional. In that case, we have

$$\mathcal{H} = -i\hbar v \partial_z + \sum_n \hbar\omega_n a_n^\dagger a_n + \sum_n (W_n(z)a_n + W_n^*(z)a_n^\dagger) \equiv \mathcal{H}_0 + V, \qquad (S1)$$

with $\mathcal{H}_0 = -i\hbar v \partial_z + \sum_n \hbar\omega_n a_n^\dagger a_n$. Let us move into the interaction picture, defining $V_I = U_0^\dagger V U_0$ with $U_0 = \exp[-vt\partial_z - it\sum_n \omega_n a_n^\dagger a_n]$. The interaction Hamiltonian in the interaction picture results in the following expression,

$$V_I = \sum_n (W_n(z+vt)a_n e^{-i\omega_n t} + W_n^*(z+vt)a_n^\dagger e^{i\omega_n t}).$$

For this result, we made use of the identity

$$e^{vt\partial_z} f(z) e^{-vt\partial_z} = f(z+vt),$$

as the operators here are just translation operators.

To proceed, it is important to understand how the interaction potential commutes with itself at later times. For example, let us calculate

$$[V_I(t), V_I(t')] = \sum_{m,n} [W_m(z+vt)a_m e^{-i\omega_m t} + W_m^*(z+vt)a_m^\dagger e^{i\omega_m t},$$

$$W_n(z+vt')a_n e^{-i\omega_n t'} + W_n^*(z+vt')a_n^\dagger e^{i\omega_n t'}].$$

A key observation is that the $W$ operators commute with themselves and their conjugates. This is because their arguments commute with each other, so if we Taylor expand both of the $W$s, they would be written in terms of operators that commute with each other. In particular, since $[z+vt, z+vt'] = 0$ (because $[z,z]=0$), it is the case for any $f, g$ that $[f(z+vt), g(z+vt')] = 0$. This property originates from the paraxial approximation. We therefore get

$$[V_I(t), V_I(t')] = \sum_n \left( W_n(z+vt) W_n^*(z+vt') e^{-i\omega_n(t-t')} - W_n(z+vt') W_n^*(z+vt) e^{i\omega_n(t-t')} \right),$$

which can be also written as

$$[V_I(t), V_I(t')] = 2i \sum_n \text{Im}\{W_n(z+vt) W_n^*(z+vt') e^{-i\omega_n(t-t')}\}.$$

While this commutator is generally non-zero, we can to see that

$$\left[ V_I(t''), [V_I(t), V_I(t')] \right] = 0.$$

The commutator $[V_I(t), V_I(t')]$ contains only electron operators. These electron operators commute with $W^*(z+vt'')$, as well as with $a, a^\dagger$, rendering the nested commutator zero. There is a closed-form solution for the unitary time-evolution operator when the nested commutator is zero. Recalling that $U = U_0 U_I$, with

$$U_I = \text{T} e^{-\frac{i}{\hbar} \int_0^t dt' V_I(t')},$$

$U_I$ can be expressed in terms of the so-called *Magnus expansion* as

$$U_I = e^\Omega = e^{\sum_{k=0}^\infty \Omega_k},$$

where the first few terms of $\Omega$ are given as

$$\Omega_1 = -\frac{i}{\hbar} \int_0^t dt_1 V_I(t_1),$$

$$\Omega_2 = \frac{1}{2}\left(-\frac{i}{\hbar}\right)^2 \int_0^t \int_0^{t_1} dt_1 dt_2 [V_I(t_1), V_I(t_2)],$$

$$\Omega_3 = \frac{1}{6}\left(-\frac{i}{\hbar}\right)^3 \int_0^t \int_0^{t_1} \int_0^{t_2} dt_1 dt_2 dt_3 \big([V_I(t_1), [V_I(t_2), V_I(t_3)]] + [V_I(t_3), [V_I(t_2), V_I(t_1)]]\big),$$

and higher order terms are given in terms of higher order commutators. It follows that if $\Omega_k = 0$ for some $k$, then $\Omega_{j>k}$ will also be zero. From the commutators we evaluated for our problem, it is evident that $\Omega_{k \geq 3} = 0$, allowing us to write the interaction picture unitary evolution operator as

$$U_I = e^{\sum_n \chi_n} e^{\sum_n \tilde{S}_n},$$

where

$$\chi_n(z,t) = -\frac{i}{\hbar^2} \int_0^t \int_0^{t_1} dt_1 dt_2 \text{Im}\{W_n(z + vt_1) W_n^*(z + vt_2) e^{-i\omega_n(t_1 - t_2)}\},$$

and

$$\tilde{S}_n(z,t) = -\frac{i}{\hbar} \int_0^t dt' \big(W_n(z + vt') a_n e^{-i\omega_n t'} + W_n^*(z + vt') a_n^\dagger e^{i\omega_n t'}\big).$$

Now we wish to simplify $\tilde{S}_n$ by defining

$$\tilde{\beta}_n(z,t) = -\frac{i}{\hbar} \int_0^t dt' W_n^*(z + vt') e^{i\omega_n t'},$$

which allows us to write $\tilde{S}_n$ in $U_I$ as

$$U_I = e^{\Sigma_n \chi_n} e^{\Sigma_n \left(\tilde{\beta}_n(z,t) a_n^\dagger - \tilde{\beta}_n^*(z,t) a_n\right)}. \tag{S2}$$

Let us examine the $\tilde{\beta}_n$ expression a bit more. Expressing $W$ in terms of more elementary quantities, we get

$$\tilde{\beta}_n = -\frac{iev}{\hbar} \int_0^t dt' A_n^*(z+vt') e^{i\omega_n t'} = \frac{ev}{\hbar \omega_n} \int_0^t dt' E_n^*(z+vt') e^{i\omega_n t'}. \tag{S3}$$

Now we are ready to make the necessary simplifications and changes needed to retrieve the equations we present in the main text.

- Looking at equation S1, we consider only a single mode case, thus removing all $n$ sums and indices. We additionally plug back $W$ in the one-dimensional case ($W = evA_z(z)$). This yields exactly equation 3 from the main text.

- Because no observables depend on our denoted start time ($t = 0$), we may choose to set our start time to minus infinity. Additionally, we are interested in what happens long after the electron has left the interaction region, i.e. $t \to \infty$. This means that the boundaries for all the time integrals are now $(-\infty, \infty)$.

- To retrieve equations 5 and 6 of the main text, we develop $\tilde{\beta}$ some more:

$$\tilde{\beta} = \frac{ev}{\hbar \omega} \int_{-\infty}^{\infty} dt E_z^*(z+vt) e^{i\omega t}.$$

We can express $E_z^*$ in terms of its Fourier transform $\hat{E}_z$ as

$$E_z^*(z) = \frac{1}{2\pi} \int_{-\infty}^{\infty} dk\, e^{ikz} \hat{E}_z(k)$$

Plugging this back in to the expression of $\tilde{\beta}$

$$\tilde{\beta}(z) = \frac{ev}{\hbar\omega} \int_{-\infty}^{\infty} dt\, e^{i\omega t} \frac{1}{2\pi} \int_{-\infty}^{\infty} dk\, e^{ikz} e^{ikvt} \hat{E}_z(k).$$

Taking first the integral over $t$

$$\tilde{\beta}(z) = \frac{ev}{\hbar\omega} \int_{-\infty}^{\infty} dk\, \delta(kv+\omega) e^{ikz} \hat{E}_z(k) = \frac{e}{\hbar\omega} e^{-i\frac{\omega}{v}z} \hat{E}_z\left(-\frac{\omega}{v}\right).$$

Plugging in the Fourier relation between $\hat{E}_z$ and $E_z$

$$\tilde{\beta}(z) = \left[\frac{e}{\hbar\omega} \int_{-\infty}^{\infty} dz'\, e^{-i\frac{\omega}{v}z'} E_z(z')\right] e^{-i\frac{\omega}{v}z} \triangleq g_{Qu} b,$$

where $b$ is the electron ladder lowering operator, for which $[b, b^\dagger] = 0$. This expression is exactly $g_{Qu}$ as defined in equation 6 of the main text. Plugging this new $\tilde{\beta}$ into $U_I$ gives us the final evolution operator, as found in equation 5 (with the addition of the now explicitly defined global phase $\chi$):

$$U(t \to \infty) = e^\chi \exp[\tilde{\beta}(z)a^\dagger - \tilde{\beta}^*(z)a] = e^\chi \exp\left[g_{Qu} e^{-i\frac{\omega}{v}z} a^\dagger - g_{Qu}^* e^{i\frac{\omega}{v}z} a\right]$$

$$\equiv e^\chi \exp[g_{Qu} b a^\dagger - g_{Qu}^* b^\dagger a]$$

### S1.2. Explicit Matrix Representation of the Scattering Operator S

It is useful to have the explicit matrix elements of the scattering operator $S$, in a closed-form expression. In this section we derive these $S$ matrix elements, which are in equations $7a$ and $7b$ of the main text. We begin from the final expression we found for $S$ in equation 5 of the main text:

$$S = \exp[g_{Qu} b a^\dagger - g_{Qu}^* b^\dagger a].$$

We want to split this exponent to a product of two. We will use the Baker-Campbell-Hausdorff formula

$$e^X e^Y = e^Z,$$

where

$$\begin{cases} X = g_{Qu} b a^\dagger \\ Y = -g_{Qu}^* b^\dagger a \\ Z = X + Y + \frac{1}{2}[X,Y] + \frac{1}{12}[X,[X,Y]] - \frac{1}{2}[Y,[X,Y]] + \cdots \end{cases}$$

First, we calculate the different commutators

$$[X,Y] = [g_{Qu} b a^\dagger, -g_{Qu}^* b^\dagger a] = -|g_{Qu}|^2 [ba^\dagger, b^\dagger a] = -|g_{Qu}|^2 (ba^\dagger b^\dagger a - b^\dagger a b a^\dagger).$$

Since $a$ and $b$ act on different particles, they commute and we get

$$[X,Y] = -|g_{Qu}|^2 (a^\dagger a b b^\dagger - a a^\dagger b^\dagger b).$$

From the commutation relation of $b$, and from the way $b, b^\dagger$ operates on electron states, we get $bb^\dagger = b^\dagger b = 1$ and so

$$[X,Y] = -|g_{Qu}|^2 (a^\dagger a - a a^\dagger) = -|g_{Qu}|^2 [a^\dagger, a] = |g_{Qu}|^2 [a, a^\dagger] = |g_{Qu}|^2,$$

where in the last equality we used the known commutator of the quantum harmonic oscillator $[a, a^\dagger] = 1$. Since the commutator we found is a constant, any higher orders will be 0. This means

$$Z = X + Y + \frac{1}{2}|g_{Qu}|^2.$$

Now, explicitly writing out the Baker-Campbell-Hausdorff formula we get

$$e^Z = \exp[g_{Qu}ba^\dagger - g_{Qu}^*b^\dagger a]\, e^{\frac{1}{2}|g_{Qu}|^2} = e^X e^Y = \exp[g_{Qu}ba^\dagger]\exp[-g_{Qu}^*b^\dagger a],$$

which finally means that

$$S = \exp[g_{Qu}ba^\dagger - g_{Qu}^*b^\dagger a] = e^{-\frac{1}{2}|g_{Qu}|^2}\exp[g_{Qu}ba^\dagger]\exp[-g_{Qu}^*b^\dagger a].$$

From here, we take the full Taylor expansion of each exponential operator to get

$$S = e^{-\frac{1}{2}|g_{Qu}|^2}\sum_{s,r=0}^{\infty}\frac{g_{Qu}^s b^s a^{\dagger s}}{s!}\frac{(-g_{Qu}^*)^r b^{\dagger r} a^r}{r!}. \tag{S4}$$

Now we are ready to calculate the matrix elements of $S$:

$$S_{k,n,k',n'} = \langle k,n|S|k',n'\rangle = \left\langle k,n\left|e^{-\frac{1}{2}|g_{Qu}|^2}\sum_{s,r=0}^{\infty}\frac{(g_{Qu})^s b^s a^{\dagger s}}{s!}\frac{(-g_{Qu}^*)^r b^{\dagger r} a^r}{r!}\right|k',n'\right\rangle$$

$$= e^{-\frac{1}{2}|g_{Qu}|^2}\sum_{s,r=0}^{\infty}\frac{g_{Qu}^s(-g_{Qu}^*)^r}{s!\,r!}\langle k,n|b^s a^{\dagger s} b^{\dagger r} a^r|k',n'\rangle.$$

Note that for any $r > n'$, we get $a^r|k',n'\rangle = 0$, which means that the $r$-sum terminates at $n'$.

Applying the operators one by one gives us:

$$S_{k,n,k',n'}$$

$$= e^{-\frac{1}{2}|g_{Qu}|^2}\sum_{s=0}^{\infty}\sum_{r=0}^{n'}\frac{(g_{Qu})^s(-g_{Qu}^*)^r}{s!\,r!}\left\langle k,n\left|1\cdot\sqrt{\frac{(n'-r+s)!}{(n'-r)!}}\cdot 1\cdot\sqrt{\frac{n'!}{(n'-r)!}}\right|k'+r-s,n'-r+s\right\rangle.$$

Using the orthonormality of the state basis, this simplifies to

$$S_{k,n,k',n'} = e^{-\frac{1}{2}|g_{Qu}|^2}\sum_{s=0}^{\infty}\sum_{r=0}^{n'}\frac{(g_{Qu})^s(-g_{Qu}^*)^r}{s!\,r!}\frac{\sqrt{(n'-r+s)!\,n'!}}{(n'-r)!}\delta_{k,k'+r-s}\delta_{n,n'-r+s}.$$

Focusing on the Kronecker delta function requirements

$$\begin{cases} k = k' + r - s \\ n = n' - r + s \end{cases} \rightarrow \begin{cases} s - r = k' - k \\ n - n' = s - r \end{cases} \rightarrow \begin{cases} k + n = k' + n' \\ s = r + (n - n') \end{cases},$$

which means that

$$\delta_{k,k'+r-s}\delta_{n,n'-r+s} = \delta_{k+n,k'+n'}\delta_{s,r+(n-n')}.$$

We start by eliminating the $s$-delta. Since $s \geq 0$, in order for the delta function to be 1, we must require $s = r + (n - n') \geq 0$, which means $r \geq n' - n$. Since $r$ must also be non-negative, the $r$-sum begins at $r = \max\{0, n' - n\}$. This results in

$$S_{k,n,k',n'} = \left[ e^{-\frac{1}{2}|g_{Qu}|^2} g_{Qu}^{n-n'} \sqrt{n! \, n'!} \sum_{r=\max\{0,n'-n\}}^{n'} \frac{\left(-|g_{Qu}|^2\right)^r}{r! \, (n' - r)! \, (r + n - n')!} \right] \delta_{k+n,k'+n'}.$$

In conclusion, we get

$$S_{k,n,k',n'} = s_{n,n'} \cdot \delta_{k+n,k'+n'},$$

where

$$s_{n,n'} = e^{-\frac{1}{2}|g_{Qu}|^2} g_{Qu}^{n-n'} \sqrt{n! \, n'!} \sum_{r=\max\{0,n'-n\}}^{n'} \frac{\left(-|g_{Qu}|^2\right)^r}{r! \, (n' - r)! \, (r + n - n')!},$$

which are exactly equations 7a,7b of the main text.

### S1.3. Analysis of a Single QPINEM Interaction

In this section, we derive equation 11 of the main text, which gives us the exact amplitude coefficients of the output state after the electron interaction with the photonic state. The QPINEM interaction provides the amplitude coefficients of the output state.

Consider the following pure input state, as in equation 1 of the main text:

$$|\psi^{(i)}\rangle = \sum_{\substack{k'=-\infty \\ n'=0}}^{\infty} c^{(i)}_{k',n'} |k', n'\rangle$$

The output state is of a similar form

$$|\psi^{(f)}\rangle = S|\psi^{(i)}\rangle = \sum_{\substack{k'=-\infty \\ n'=0}}^{\infty} c^{(i)}_{k',n'} S|k', n'\rangle \triangleq \sum_{\substack{k=-\infty \\ n=0}}^{\infty} c^{(f)}_{k,n} |k, n\rangle$$

The output amplitude coefficient $c^{(f)}_{k,n}$ can be retrieved by

$$c^{(f)}_{k,n} = \langle k, n|\psi^{(f)}\rangle = \sum_{\substack{k'=-\infty \\ n'=0}}^{\infty} c^{(i)}_{k',n'} \langle k, n|S|k', n'\rangle$$

These are exactly the matrix elements we have already found in the previous section. Thus

$$c^{(f)}_{k,n} = \sum_{\substack{k'=-\infty \\ n'=0}}^{\infty} c^{(i)}_{k',n'} S_{k,n,k',n'} = \sum_{\substack{k'=-\infty \\ n'=0}}^{\infty} c^{(i)}_{k',n'} S_{n,n'} \delta_{k+n,k'+n'} = \sum_{\substack{k'=-\infty \\ n'=0}}^{\infty} c^{(i)}_{k',n'} S_{n,n'} \delta_{k+n-n',k'}$$

Since $k'$ ranges from minus infinity to infinity, there always exists a $k'$ such that $k' = k + n - n'$ for any $k, n, n'$. Therefore

$$c^{(f)}_{k,n} = \sum_{n'=0}^{\infty} c^{(i)}_{k+n-n',n'} S_{n,n'},$$

which is exactly equation 11 of the main text.

Knowing the amplitude coefficients allows us to calculate the output density matrix, and may contain more information than the density matrix alone (e.g. in the case of multiple consecutive

QPINEM interactions that involve tracing-out of the state). In other words, we can always find the density matrix from a given set of amplitude coefficients, but we cannot always go in the opposite direction (as in the case of mixed states).

## S2. Assisting Derivations for Section II: Electron Interaction with a Coherent State

The common setup in PINEM experiments and theory is that of a coherent photonic state and a baseline-energy electron state. In this section, we explicitly derive two expression for the resulting photonic state of this interaction, which relate to equations 12 and 13 of the main text. Recall the explicit output amplitude coefficients formula, given by equation 11:

$$c_{k,n}^{(f)} = \sum_{n'=0}^{\infty} c_{k+n-n',n'}^{(i)} S_{n,n'}$$

$$= e^{-\frac{1}{2}|g_{Qu}|^2} \sum_{n'=0}^{\infty} c_{k+n-n',n'}^{(i)} g_{Qu}^{n-n'} \sqrt{n!\, n'!} \sum_{r=\max\{0,n'-n\}}^{n'} \frac{\left(-|g_{Qu}|^2\right)^r}{r!\,(n'-r)!\,(r+n-n')!}.$$

A coherent photon state has the following wavefunction:

$$|\alpha\rangle_p = e^{-\frac{1}{2}|\alpha|^2} \sum_{n=0}^{\infty} \frac{\alpha^n}{\sqrt{n!}} |n\rangle.$$

Together with a "*delta*" electron $|\delta\rangle_e \triangleq |0\rangle_e$, the input state is

$$|\psi^{(i)}\rangle = e^{-\frac{1}{2}|\alpha|^2} \sum_{n=0}^{\infty} \frac{\alpha^n}{\sqrt{n!}} |0,n\rangle = \sum_{\substack{k=-\infty \\ n=0}}^{\infty} \left(\delta_{k,0} \cdot e^{-\frac{1}{2}|\alpha|^2} \frac{\alpha^n}{\sqrt{n!}}\right) |k,n\rangle.$$

This means that our input amplitude coefficients are

$$c_{k,n}^{(i)} = \delta_{k,0} \cdot e^{-\frac{1}{2}|\alpha|^2} \frac{\alpha^n}{\sqrt{n!}}.$$

Plugging this result into $c_{k,n}^{(f)}$:

$$c_{k,n}^{(f)}$$

$$= e^{-\frac{1}{2}|g_{Qu}|^2} \sum_{n'=0}^{\infty} \left(\delta_{k+n-n',0} e^{-\frac{1}{2}|\alpha|^2} \frac{\alpha^{n'}}{\sqrt{n'!}}\right) g_{Qu}^{n-n'} \sqrt{n! n'!} \sum_{r=\max\{0,n'-n\}}^{n'} \frac{\left(-|g_{Qu}|^2\right)^r}{r!(n'-r)!(r+n-n')!}$$

$$= e^{-\frac{|g_{Qu}|^2+|\alpha|^2}{2}} \sum_{n'=0}^{\infty} \left(\delta_{k+n,n'} \alpha^{n'}\right) g_{Qu}^{n-n'} \sqrt{n!} \sum_{r=\max\{0,n'-n\}}^{n'} \frac{\left(-|g_{Qu}|^2\right)^r}{r!(n'-r)!(r+n-n')!}$$

Note that for the delta function to equal one, we need to sum over some $n'$ such that $n' = k + n$. Since $n' > 0$, this immediately gives us that $c_{k,n}^{(f)}$ is always 0 for any $k + n < 0$. For the complementary case of $k + n \geq 0$, we get

$$c_{k,n}^{(f)} = e^{-\frac{|g_{Qu}|^2+|\alpha|^2}{2}} \alpha^{k+n} g_{Qu}^{-k} \sqrt{n!} \sum_{r=\max\{0,k\}}^{k+n} \frac{\left(-|g_{Qu}|^2\right)^r}{r!(k+n-r)!(r-k)!}.$$

Exactly as shown in equation 12 of the main text.

To derive equation 13, we go back to the Taylor expansion representation of $S$, as given by (S4). Applying it to our input system state gives

$$|\psi^{(f)}\rangle = S|\psi^{(i)}\rangle = S(|\delta\rangle_e \otimes |\alpha\rangle_p) = e^{-\frac{1}{2}|g_{Qu}|^2} \sum_{s,r=0}^{\infty} \frac{g_{Qu}^s(-g_{Qu}^*)^r}{s!r!} [b^s b^{\dagger r}|0\rangle_e] \otimes [a^{\dagger s} a^r |\alpha\rangle_p].$$

Since photonic coherent states are eigenstates of the annihilation operator, we get $a^r|\alpha\rangle_p = \alpha^r|\alpha\rangle_p$ and

$$|\psi^{(f)}\rangle = e^{-\frac{1}{2}|g_{Qu}|^2} \sum_{s,r=0}^{\infty} \frac{g_{Qu}^s(-g_{Qu}^*\alpha)^r}{s!\,r!} [b^s b^{\dagger r}|0\rangle_e] \otimes [a^{\dagger s}|\alpha\rangle_p].$$

Applying the electron ladder operators to the electron state results in

$$|\psi^{(f)}\rangle = e^{-\frac{1}{2}|g_{Qu}|^2} \sum_{s,r=0}^{\infty} \frac{g_{Qu}^s(-g_{Qu}^*\alpha)^r}{s!\,r!} |r-s\rangle_e \otimes [a^{\dagger s}|\alpha\rangle_p].$$

Post-selecting an electron of an energy state $k$ is equivalent to taking the state $\langle k|S|\psi^{(i)}\rangle$. This means that the photonic output state is

$$|\psi^{(f)}\rangle_p = \langle k|S|\psi^{(i)}\rangle = e^{-\frac{1}{2}|g_{Qu}|^2} \sum_{s,r=0}^{\infty} \frac{g_{Qu}^s(-g_{Qu}^*\alpha)^r}{s!\,r!} \langle k|r-s\rangle_e [a^{\dagger s}|\alpha\rangle_p]$$

$$= e^{-\frac{1}{2}|g_{Qu}|^2} \sum_{s,r=0}^{\infty} \frac{g_{Qu}^s(-g_{Qu}^*\alpha)^r}{s!\,r!} \delta_{k,r-s} [a^{\dagger s}|\alpha\rangle_p]$$

$$= e^{-\frac{1}{2}|g_{Qu}|^2} \sum_{s,r=0}^{\infty} \frac{g_{Qu}^s(-g_{Qu}^*\alpha)^r}{s!\,r!} \delta_{s,r-k} [a^{\dagger s}|\alpha\rangle_p]$$

For the case of an electron energy loss ($k \leq 0$), we always sum on some $s$ such that $s = r - k \geq r$, and so the delta is not always zero. However, if $k > 0$ then there exists an $s \geq 0$ such that $s = r - k$ if and only if $r \geq k$. This means the $r$-sum begins at $\max\{0, k\}$. We thus get

$$|\psi^{(f)}\rangle_p = e^{-\frac{1}{2}|g_{Qu}|^2} \sum_{r=\max\{0,k\}}^{\infty} \frac{g_{Qu}^{r-k}(-g_{Qu}^*\alpha)^r}{(r-k)!\,r!} [(a^\dagger)^{r-k}|\alpha\rangle_p].$$

We substitute $r \to r + k$ for the following cleaner expression:

$$|\psi^{(f)}\rangle_p = e^{-\frac{1}{2}|g_{Qu}|^2}(-g_{Qu}^*\alpha)^k \sum_{r=\max\{-k,0\}}^{\infty} \frac{(-\alpha|g_{Qu}|^2)^r}{r!\,(r+k)!} [(a^\dagger)^r|\alpha\rangle_p].$$

This result is exactly equation 13 of the main text, although now we see the explicit normalization needed for the state. For the sake of the qualitative behavior, this normalization is not important, and hence was left out of the main text.

## S3. Assisting Derivations for Section III: Creation of Photonic Fock States

In section III of the main text, we present a scheme to generate a photonic Fock state, starting with an empty cavity, and using multiple QPINEM interactions with "*delta*" electrons. In this section, we begin by proving that the scheme in figure (3a) of the main text always results in a photonic Fock state, after every interaction. We prove this claim using induction on the interaction number $m$.

- For $m = 0$, we begin with an empty cavity, $|\psi^{(i)}\rangle_\mathrm{p} = |0\rangle_\mathrm{p}$, which is the "zeroth" Fock state.
- We assume that after interaction $m$, and the measurement of the $m$'th electron, we have some photonic Fock state $|\psi^{(f,m)}\rangle_\mathrm{p} = |N_m\rangle_\mathrm{p}$.
- After introducing a new "*delta*" electron, our input system state to the $(m+1)$st interaction is

$$|\psi^{(i,m+1)}\rangle = |0, N_m\rangle$$

    Using the same claim from the main text, $k + n$ is conserved before and after the interaction, per input basis state. Since our whole system state is a single basis state, then $k + n$ after the interaction must also be $N_m$. Assuming we measured some electron energy

$k'$, then the only possible photonic state is $|n' = N_m - k'\rangle_p$, a photonic Fock state, thus concluding our proof.

Additionally, for this section, we explicitly use equation 11 to find the photonic state after each interaction. Since, however, this process is stochastic and dependent on the measured electron energies, we assume we know the photonic Fock state at the end of some interaction numbered $m$. Our input system state to interaction number $m + 1$ is

$$|\psi^{(i,m+1)}\rangle = |0, N_m\rangle = \sum_{\substack{k=-\infty \\ n=0}}^{\infty} \delta_{k,0}\delta_{n,N_m}|k, n\rangle,$$

which means that

$$c_{k,n}^{(i,m+1)} = \delta_{k,0}\delta_{n,N_m}.$$

Plugging this into equation 11 yields

$$c_{k,n}^{(f,m+1)} = \sum_{n'=0}^{\infty} c_{k+n-n',n'}^{(i,m+1)} S_{n,n'}$$

$$= e^{-\frac{1}{2}|g_{Qu}|^2} \sum_{n'=0}^{\infty} \delta_{k+n-n',0}\delta_{n',N_m} g_{Qu}^{n-n'} \sqrt{n!\, n'!} \sum_{r=\max\{0,n'-n\}}^{n'} \frac{\left(-|g_{Qu}|^2\right)^r}{r!\,(n'-r)!\,(r+n-n')!}.$$

Similarly to section S2, we find that $k + n$ must be nonnegative for the sum to be nonzero (because of the delta). For $k + n \geq 0$, we get

$$c_{k,n}^{(f,m+1)} = e^{-\frac{1}{2}|g_{Qu}|^2} g_{Qu}^{-k} \sqrt{n!\,(n+k)!} \sum_{r=\max\{0,k\}}^{n+k} \frac{\left(-|g_{Qu}|^2\right)^r}{r!\,(n+k-r)!\,(r-k)!} \delta_{n+k,N_m}.$$

The delta function $\delta_{n+k,N_m}$ limits the range of summation by cutting the cases $k+n < 0$, because $k+n = N_m \geq 0$, with $N_m$ representing a Fock-state number. Expressing the amplitude coefficients strictly with the Fock number $n$, we get

$$c_{k,n}^{(\mathrm{f},m+1)} = e^{-\frac{1}{2}|g_{\mathrm{Qu}}|^2} g_{\mathrm{Qu}}{}^{n-N_m} \sqrt{n!\, N_m!} \sum_{r=\max\{0,N_m-n\}}^{N_m} \frac{\left(-|g_{\mathrm{Qu}}|^2\right)^r}{r!\,(N_m-r)!\,(r-N_m+n)!},$$

which finally reveals that the probability to transition to a Fock state $|n\rangle_\mathrm{p}$, assuming we were at a Fock state $|N_m\rangle_\mathrm{p}$ is

$$P(N_m \to n) = \left|c_n^{(\mathrm{f},m+1)}\right|^2$$

$$= e^{-|g_{\mathrm{Qu}}|^2} |g_{\mathrm{Qu}}|^{2(n-N_m)} n!\, N_m! \left|\sum_{r=\max\{0,N_m-n\}}^{N_m} \frac{\left(-|g_{\mathrm{Qu}}|^2\right)^r}{r!\,(N_m-r)!\,(r-N_m+n)!}\right|^2. \quad (S5)$$

## S4. Assisting Derivations for Section V: Displacement of Photonic Coherent States

In section V of the main text, we presented a new useful electron state, a "*comb*" electron, as well as demonstrated its use in a scheme to displace photonic coherent states. In this section, we begin by proving that the electron "*comb*" state in equation 14 of the main text is indeed an eigenvector of $b$ with an eigenvalue $\beta$. Recall

$$|c\rangle_\mathrm{e} = \lim_{K,K' \to \infty} \frac{1}{\sqrt{K+K'+1}} \sum_{k=-K}^{K'} \beta^k |k\rangle_\mathrm{e}.$$

Applying $b$ to our state

$$b|c\rangle_e = \lim_{K,K'\to\infty} \frac{1}{\sqrt{K+K'+1}} \sum_{k=-K}^{K'} \beta^k b|k\rangle_e = \lim_{K,K'\to\infty} \frac{1}{\sqrt{K+K'+1}} \sum_{k=-K}^{K'} \beta^k |k-1\rangle_e.$$

Performing a variable change $k \to k+1$

$$b|c\rangle_e = \lim_{K,K'\to\infty} \frac{1}{\sqrt{K+K'+1}} \sum_{k=-K-1}^{K'-1} \beta^{k+1} |k\rangle_e = \beta \left[ \lim_{K,K'\to\infty} \frac{1}{\sqrt{K+K'+1}} \sum_{k=-K-1}^{K'-1} \beta^k |k\rangle_e \right].$$

Noticing the resemblance of the sum to the original comb state

$$b|c\rangle_e = \beta|c\rangle_e + \beta \cdot \lim_{K,K'\to\infty} \frac{-\beta^{K'}|K'\rangle_e + \beta^{-K-1}|-K-1\rangle_e}{\sqrt{K+K'+1}}.$$

We now require $|\beta| = 1$, as the coefficient $\beta^k/\sqrt{K+K'+1}$ in $|c\rangle_e$ will diverge otherwise, and the electron state will not be normalizable. In that case, the limit above tends to zero and we indeed get that

$$b|c\rangle_e = \beta|c\rangle_e.$$

One can use an identical approach to prove that $|c\rangle_e$ is in fact an eigenvector of $b^\dagger$ as well, with an eigenvalue $\beta^*$.

Next, we prove the operator equivalency in equation 15 of the main text. That is, we prove that when interacting with a "*comb*" electron, the scattering operator is equivalent to a scalar displacement. This result is shown by using again the Taylor expansion of $S$ from equation S4. Applying this equation to our system state, the final state is

$$|\psi^{(f)}\rangle = S|\psi^{(i)}\rangle = e^{-\frac{1}{2}|g_{Qu}|^2} \sum_{s,r=0}^{\infty} \frac{g_{Qu}^s b^s a^{\dagger s}}{s!} \frac{(-g_{Qu}^*)^r b^{\dagger r} a^r}{r!} |\alpha\rangle_p \otimes |c\rangle_e.$$

Remembering that operators of different particles commute, we get

$$|\psi^{(f)}\rangle = e^{-\frac{1}{2}|g_{Qu}|^2} \sum_{s,r=0}^{\infty} \frac{g_{Qu}^s a^{\dagger s}}{s!} \frac{(-g_{Qu}^*)^r a^r}{r!} |\alpha\rangle_p \otimes [b^s b^{\dagger r} |c\rangle_e]$$

$$= e^{-\frac{1}{2}|g_{Qu}|^2} \sum_{s,r=0}^{\infty} \frac{g_{Qu}^s a^{\dagger s}}{s!} \frac{(-g_{Qu}^*)^r a^r}{r!} |\alpha\rangle_p \otimes [\beta^s (\beta^*)^r |c\rangle_e]$$

$$= e^{-\frac{1}{2}|g_{Qu}|^2} \sum_{s,r=0}^{\infty} \frac{(\beta g_{Qu})^s a^{\dagger s}}{s!} \frac{((-\beta g_{Qu})^*)^r a^r}{r!} |\alpha\rangle_p \otimes |c\rangle_e.$$

We retract back the Taylor expansion and find

$$|\psi^{(f)}\rangle = D(\beta g_{Qu})|\psi^{(i)}\rangle.$$

Thus proving the equivalency: $S = D(b g_{Qu}) \Leftrightarrow D(\beta g_{Qu})$.

Lastly, we explicitly prove the photonic state evolution of equation 16 of the main text, i.e. the displacement of the photonic coherent state. For any two complex parameters, the displacement operator satisfies

$$D(\alpha_1)D(\alpha_2) = \exp\left[\frac{\alpha_1 \alpha_2^* - \alpha_1^* \alpha_2}{2}\right] D(\alpha_1 + \alpha_2).$$

Since scalar displacement only acts on the photonic state, the output state is given by

$$|\psi^{(f)}\rangle = D(\beta g_{Qu})|\psi^{(i)}\rangle = [D(\beta g_{Qu})|\alpha\rangle_p] \otimes |c\rangle_e = [D(\beta g_{Qu})D(\alpha)|0\rangle_p] \otimes |c\rangle_e.$$

Adding up the two displacements, using the mentioned property of $D$

$$|\psi^{(f)}\rangle = \exp\left[\frac{\beta g_{\text{Qu}}\alpha^* - \beta^* g_{\text{Qu}}^*\alpha}{2}\right][D(\alpha + \beta g_{\text{Qu}})|0\rangle_p]\otimes|c\rangle_e$$
$$= \exp[-i\cdot\text{Im}\{\alpha\beta^* g_{\text{Qu}}^*\}]\,|\tilde{\alpha} = \alpha + \beta g_{\text{Qu}}\rangle_p\otimes|c\rangle_e. \tag{S6}$$

Indeed, we found the relation of equation 16. We also got the exact global phase that is added every interaction. However, this global phase has no effect on any observables, and so was dropped from the main text. From this point, it is possible to generalize this process using induction for any number of interactions $M$, and for any sequence of $\{\beta_m, g_{\text{Qu},m}\}_m$ (which may vary between interactions).

## S5. Assisting Derivations for Section VI: Creation of Displaced Fock States

In section VI of the main text, we have presented a scheme for the generation of photonic displaced Fock states, starting from some photonic Fock state, and using "*comb*" electrons. The state evolution in equation 17 of the main text can be proven in the exact same manner as the derivation in the previous section. Note that this time we do not get any global phase, because all of the displacements that we add up have the exact same argument. In that case

$$D(\alpha)D(\alpha) = D(2\alpha).$$

We do show, however, how the expression for the output photonic state is retrieved, containing an explicit sum of photon-added coherent states, in equation 18. Let us consider a displaced Fock state $|N_i, \alpha\rangle$. We may express it as

$$|N_i, \alpha\rangle_p = D(\alpha) \frac{a^{\dagger N_i}}{\sqrt{N_i}} |0\rangle_p.$$

Using the two relations

$$\begin{cases} D^{-1}(\alpha) = D(-\alpha) \\ D(\alpha) a^\dagger D(-\alpha) = a^\dagger - \alpha' \end{cases}$$

We get

$$|N_i, \alpha\rangle_p = D(\alpha) \frac{a^{\dagger N_i}}{\sqrt{N_i}} D(-\alpha) D(\alpha) |0\rangle_p = \frac{1}{\sqrt{N_i}} (a^\dagger - \alpha)^{N_i} D(\alpha) |0\rangle_p.$$

Using once again the Baker-Campbell-Hausdorff formula (though this time the commutators are much simpler, because only a single operator is present)

$$|N_i, \alpha\rangle_p = e^{-\frac{1}{2}|\alpha|^2} \frac{1}{\sqrt{N_i}} (a^\dagger - \alpha)^{N_i} \exp[\alpha a^\dagger] |0\rangle_p = e^{-\frac{1}{2}|\alpha|^2} \frac{1}{\sqrt{N_i}} \left(\frac{d}{d\alpha} - \alpha\right)^{N_i} \exp[\alpha a^\dagger] |0\rangle_p.$$

Using the binomial expansion

$$|N_i, \alpha\rangle_p = e^{-\frac{1}{2}|\alpha|^2} \frac{1}{\sqrt{N_i}} \sum_{r=0}^n \frac{n!}{r!(n-r)!} (-\alpha)^{n-r} \frac{d^r}{d\alpha^r} \exp[\alpha a^\dagger] |0\rangle_p$$

$$= e^{-\frac{1}{2}|\alpha|^2} \frac{1}{\sqrt{N_i}} \sum_{r=0}^n \frac{n!}{r!(n-r)!} (-\alpha)^{n-r} a^{\dagger r} \exp[\alpha a^\dagger] |0\rangle_p.$$

We retract back the displacement operator, again using the Baker-Campbell-Hausdorff formula, and get

$$|N_i, \alpha\rangle_p = \frac{1}{\sqrt{N_i}} \sum_{r=0}^{n} \frac{n!}{r!(n-r)!} (-\alpha)^{n-r} a^{\dagger r} D(\alpha)|0\rangle_p$$

$$= \frac{1}{\sqrt{N_i}} \sum_{r=0}^{n} (-\alpha)^{n-r} \frac{n!}{r!(n-r)!} [a^{\dagger r}|\alpha\rangle_p]. \qquad (S7)$$

This result is precisely equation 18 of the main text, with the addition of the explicit normalization (the normalization was dropped in the main text because it has no effect on the qualitative behavior).